\documentstyle[aps,prd,floats,epsf,amsfonts]{revtex}

\newcommand{\ph}{{\phantom{\dagger}}}
\newcommand{\beq}{\begin{equation}}
\newcommand{\eeq}{\end{equation}}
\newcommand{\bea}{\begin{eqnarray}}
\newcommand{\eea}{\end{eqnarray}}
\newcommand{\nn}{\nonumber}

\begin{document}
\draft
\title{Flow equation approach to the sine--Gordon model\footnote{This work
is dedicated to Prof.~Wegner on the occasion of his 60th birthday.}}
\author{Stefan Kehrein \cite{leave}}
\address{Lyman Laboratory of Physics, Harvard University, Cambridge, MA 02138}

\date{June 23, 2000}

\maketitle

\begin{abstract}
A continuous sequence of infinitesimal unitary transformations is used to diagonalize 
the \linebreak quantum sine--Gordon model for $\beta^2\in(2\pi,\infty)$. This approach can be understood
as an extension of perturbative scaling theory since it links weak-- to strong--coupling
behavior in a systematic expansion: a small expansion parameter is identified and
this parameter remains small throughout the entire flow unlike the diverging running coupling
constant of perturbative scaling. Our approximation consists in neglecting higher orders in 
this small parameter. We find very accurate results for the single--particle/hole spectrum 
in the strong--coupling phase and can describe the full crossover from weak to strong--coupling. 
The integrable structure
of the sine--Gordon model is not used in our approach. Our new method
should be of interest for the investigation of nonintegrable perturbations and 
for other strong--coupling problems. 
\end{abstract}


\section{Introduction} 
\subsection{Motivation} \noindent
Perturbative scaling theory plays a key role for analyzing the large
class of physical systems with a mismatch between the high--energy
scale of the model and the experimentally interesting low--energy scale.
For example, in field theory one is generally interested in the universal
properties at energies much lower than the UV~(ultraviolet)--cutoff, or
in condensed matter physics in energies and temperatures much smaller
than the Fermi energy/temperature usually of order of a few 1000~K. In
order to link high--energy and low--energy regimes it is of fundamental
importance to perform perturbation theory in a {\em stable} order by
first analyzing the effect from high--energy scales, and then progressively
smaller energies. An elementary example for this procedure is provided
by atomic physics, where one e.g.\ first 
establishes the fine structure of a spectrum before using these states to 
evaluate the hyperfine splittings.

For systems with continuous energy scales, like in field theory, the
above observations have led to the development of {\em perturbative
scaling theory}. Perturbative scaling ideas have become a key theoretical tool for
analyzing physical systems with many degrees of freedom. The principal
idea is to study {\em perturbatively} the effect of lowering the
high--energy cutoff by finding a Hamiltonian with this reduced cutoff
and renormalized couplings that describe 
the same low--energy physics as the original Hamiltonian. In a path
integral formulation this is conveniently achieved by successively integrating 
out the high--energy degrees of freedom. 

This procedure leads to the well--known renormalization group (RG)
equations that describe the flow of the running coupling
constants upon lowering the UV--cutoff. For the important class of
{\em strong--coupling} problems, however, the RG--equations lead 
to running coupling constants that grow larger and larger at smaller
energy scales (and often eventually even diverge). Since the RG--equations
themselves are derived perturbatively, this means that the 
perturbative scaling approach breaks down for strong--coupling problems.
Well--known examples for this class of models are the Kondo model
in condensed matter physics or QCD in elementary particle physics. 
In spite of its eventual breakdown, perturbative scaling can
still contribute significantly to the understanding of strong--coupling problems.
For example, in the Kondo model, the divergence of the running coupling
constant occurs at an energy that sets the low--energy
Kondo scale of the model, which already allows considerable insight into
the problem. Still the approach becomes uncontrolled since the coupling
constants grow very large, and it has, so far, not been possible to extend
the perturbative scaling approach in such a way that a controlled 
systematic expansion emerges that links weak-- to strong--coupling behavior. 
One can sum up these observations by noting that the perturbative
scaling approach often allows us to identify the relevant low--energy scale of a
strong--coupling problem, but frequently not the physical behavior associated
with this energy scale or the {\em crossover} behavior linking high
and low energies. For an excellent review of these issues see Ref.~\cite{Wilson75}.

This paper will exemplify the way in which Wegner's method of {\em flow
equations} \cite{Wegner94} can overcome these shortcomings and provide
an analytic description for a weak-- to strong--coupling behavior crossover. 
In the flow equation approach, a continuous sequence of infinitesimal
unitary transformations is applied to a many--particle Hamiltonian such
that the Hamiltonian becomes successively more diagonal. Wegner has set up
this approach in a differential formulation
\beq
\frac{dH}{dB}=[\eta(B),H(B)] \ .
\eeq
Here $\eta(B)=-\eta(B)^\dagger$ is an anti--Hermitian 
operator. Therefore $H(B)$ as obtained by the solution of this differential
equation describes a one--parameter family of unitarily equivalent 
Hamiltonians. $H(B=0)=H$ is the initial condition relating us to the
original Hamiltonian~$H$ in which we are interested. {\em We want} $H(B=\infty)$ 
{\em to be diagonal.} In order to achieve this, Wegner has proposed 
a suitable choice for the generator $\eta(B)$ that we will discuss
in more detail in Sect.~III.A. Wegner's construction of $\eta$ generates a 
Hamiltonian flow where the
interaction matrix elements that couple degrees of freedoms with 
a large energy difference are removed first (for smaller~$B$), and more degenerate
matrix elements during later stages of the flow. This {\em separation
of energy scales} is reminiscent of the perturbative scaling approach
and allows a stable sequence of approximations. As opposed to the perturbative
scaling approach, however, degrees of freedom are {\em not} integrated out in
the flow equation approach, instead they are successively
diagonalized. A similar framework that contains Wegner's flow equations
as a special case has independently been developed by G{\l}azek and Wilson 
({\em similarity renormalization scheme}) \cite{GlazekWilson}.

So far the flow equation approach has been applied to a variety of models
in condensed matter theory like the $n$--orbital model \cite{Wegner94,KabelWegner97}, 
impurity models like the spin--boson model \cite{KehreinMielke97}
and the Anderson impurity model \cite{KehreinMielke96}, electron--phonon 
systems \cite{Mielke97,RagwitzWegner99} and spin models \cite{Stein00,KnetterUhrig00} 
etc.\ (for an overview see also Ref.~\cite{Wegner97}). One advantage of this scheme
lies in the observation that it is a non--perturbative approach due to the 
separation of energy scales, but still has access to all energy scales since
no degrees of freedom are integrated out. Therefore one can investigate 
correlation functions on all energy scales \cite{KehreinMielke97}. Also the
flow equation approach allows the systematic derivation of low--energy 
effective Hamiltonians not plagued with singular interactions that frequently
occur in other approaches \cite{KehreinMielke96,Mielke97}. 

However, these applications did not deal with strong--coupling problems
as defined above, which would be a very interesting perspective for this
new method. G{\l}azek and Wilson undertook a first step in this direction
in Ref.~\cite{GlazekWilson98}. They investigated
a quadratic Hamiltonian that shows strong--coupling behavior due 
to the formation of a bound state from a continuum, and
demonstrated how this model can be solved using infinitesimal unitary
transformations. However, since they dealt with a quadratic Hamiltonian,
this was not a true many--particle strong--coupling problem as would
be of most interest in condensed matter theory or high energy theory.

Recently, I described the application of the flow equation method
to the one--dimensional quantum sine--Gordon model~\cite{Kehrein99}. The
sine--Gordon model is a many--particle problem with an interesting
phase structure including a strong--coupling regime.
It was shown in Ref.\cite{Kehrein99} that it is possible to use the
flow equation scheme to develop a systematic expansion that links
weak-- to strong--coupling behavior in a controlled way. Already
the leading order of this expansion was is close
agreement with exact results. In the
present paper I will present the various details of the calculation
not included in the original Letter \cite{Kehrein99} in a self--contained
manner.

The sine--Gordon model is defined by the Hamiltonian
\begin{equation}
H=\int dx\, \left( \frac{1}{2}\Pi^2(x)+\frac{1}{2}\left(\frac{\partial\phi}{\partial x}\right)^2
+u \tau^{-2} \cos\left[\beta\phi(x)\right] \right) \ ,
\end{equation}
where $\phi(x)$ is a bosonic field and $\Pi(x)$ its conjugate momentum field with 
the commutator $[\Pi(x),\phi(x')]=-i\delta(x-x')$. $u>0$ is a small dimensionless
coupling constant and $\Lambda\propto\tau^{-1}$ an implicit UV--cutoff. 
We are interested in the universal properties for energies $|E|\ll \Lambda$.

The sine--Gordon model exhibits a strong--coupling phase for $\beta^2\lesssim 8\pi$
with a mass gap and {\em fermionic} low--energy excitations 
({\em massive solitons}). The perturbative
scaling analysis leads to a characteristic {\em strong--coupling divergence}
of the running coupling~$u$ in this regime. This makes the sine--Gordon model an 
interesting test model for our new approach. The main emphasis in this paper will 
not lie in deriving new results in this well--studied model, 
but in showing how these results follow within the flow equation method, and how
therefore our new method can be useful for strong--coupling problems more
generally.

Other features that make the sine--Gordon model an attractive test model
are its interesting phase structure with a Kosterlitz--Thouless type transition
to a phase with massless solitons at $\beta^2/8\pi \approx 1+O(u)$, its 
integrable structure that allows 
the comparison with exact results \cite{Faddeev75,Zamolodchikov77}, 
and its relation to a variety of other models like the spin-$1/2$ X-Y-Z chain, 
the $1d$~electron gas with backward scattering, the Thirring model in field theory
and the $2d$~Coulomb gas
(for an overview of these relations see Ref.~\cite{Solyom79}). Therefore
the results from the flow equation approach can be viewed within a variety
of model contexts.

The main motivation for being interested in the flow equation approach to
this integrable model lies, however, in the observation that our new method
does {\em not} make use of the integrable structure. In our approach a small
parameter is identified and used within a suitably renormalized perturbation
expansion. The usual perturbative scaling approach fails because the
initially small expansion parameter~$u$ diverges during the RG--procedure. 
In the flow equation approach the expansion parameter will turn out to
be the product of the running coupling~$u$ and a factor $(-1+\beta^2/4\pi)$.
This combination will {\em always} remain small during the flow. It is therefore
feasible to study for example nonintegrable perturbations and correlation functions
within our new approach, which should be of considerable interest in a variety
of contexts. Although the calculations presented here appear rather lengthy and
technical at first, they are straightforward and much closer to conventional 
many--body techniques than methods building on the integrable structure.

\subsection{Outline} \noindent
The structure of this paper is as follows. Sect.~II deals with some general
properties of the sine--Gordon model that are important in the sequel. In
Sect.~II.A the sine--Gordon model and the regularization used in this paper
are introduced. Sect.~II.B reviews the perturbative scaling 
analysis, the phase structure, and the strong--coupling behavior. In Sect.~II.C
various exact results based on the integrable structure of the sine--Gordon 
model are summed up, especially properties of the point $\beta^2=4\pi$ where 
the model becomes equivalent to a noninteracting Thirring model. This equivalence
will play an important role in understanding the structure of our flow equation 
approach later on. 

After setting the stage in Sect.~II, Sect.~III deals with the actual application
of the flow equation approach to the sine--Gordon model. Some general properties of the 
flow equation method are reviewed in~III.A. Then the appropriate generator~$\eta(B)$
for the sine--Gordon model diagonalization is worked out in~III.B and the 
commutator $[\eta(B),H_0]$ evaluated in~III.C. The key computational parts of
the flow equation approach are contained in III.D and III.E, where the commutators
$[\eta(B),H_{\rm int}(B)]$ and $[\eta(B),H_{\rm diag}(B)]$ are evaluated. 
From these commutators the flow of~$\beta^2(B)$ and of the running coupling constant
are deduced in Sect.~III.F.

For the convenience of the reader, all the results from this technical
part are summed up in Sect.~IV.A, in particular the Hamiltonian~$H(B)$ along
the flow and the set of flow equations governing the various parameters 
in~$H(B)$. We will see in IV.B that in the strong--coupling phase $H(B)$ flows to an
effective low--energy noninteracting Thirring model.
The mass gap of the sine--Gordon model can be easily deduced from this low--energy
model, and the results are then compared with perturbative scaling analysis and exact
integrable model results. The agreement will turn out to be very good. 
In Sect.~IV.C the final diagonal Hamiltonian $H(B=\infty)$ is discussed in more
detail, in particular the soliton dispersion relation and properties in the
crossover region. Finally in Sect.~IV.D the approximations and the 
expansion parameter of our approach are reviewed.

Sect.~V sums up the conclusions and an outlook to open questions. 
The Appendix contains important properties of vertex operators that are used
throughout this paper.

\section{Sine--Gordon Model} 
\subsection{Definition} \noindent
The one--dimensional quantum sine--Gordon model is defined by the Hamiltonian
\begin{equation}
H=\int dx\, \left( \frac{1}{2}\Pi^2(x)+\frac{1}{2}\left(\frac{\partial\phi}{\partial x}\right)^2
+u \tau^{-2} \cos\left[\beta\phi(x)\right] \right) \ .
\label{sinegordon1}
\end{equation}
$\phi(x)$ is a bosonic field and $\Pi(x)$ its conjugate momentum field with 
the fundamental commutator
\begin{equation}
[\Pi(x),\phi(x')]=-i\delta(x-x') \ .
\end{equation}
In (\ref{sinegordon1}) an UV--momentum cutoff $\Lambda\propto\tau^{-1}$ is implied.
$u$ is a dimensionless coupling constant. Without loss of generality we will assume
$u>0$ and $\beta>0$. 

Expanding the fields in normal modes gives 
\begin{eqnarray}
\phi(x)&=&-\frac{i}{\sqrt{4\pi}} {\sum_{k\neq 0}} \,\frac{\sqrt{|k|}}{k}\, e^{-ikx}
\left(\sigma_1(k)+\sigma_2(k)\right) \\
\Pi(x)&=&\frac{1}{\sqrt{4\pi}}{\sum_{k\neq 0}}\,\sqrt{|k|}\, e^{-ikx}
\left(\sigma_1(k)-\sigma_2(k)\right) \ . 
\end{eqnarray}
Sums over wavevectors $k,p,q,\ldots$ are to be understood in the sense
\begin{equation}
{\sum_k}\stackrel{\rm def}{=}\frac{2\pi}{L}\sum_{n=-\infty}^\infty
\label{wavevectorsum}
\end{equation}
with $k=2\pi n/L$ throughout this paper. $L$ is the system size. The basic commutators
for $k,k'>0$ are
\bea
[\sigma_1(-k),\sigma_1(k')]=[\sigma_2(k),\sigma_2(-k')]&=&\delta_{kk'}\, L/2\pi \nn \\
{[\sigma_j(k),\sigma_j(k')]}={[\sigma_j(-k),\sigma_j(-k')]}&=&0  
\label{comm1}
\eea
and for $j\neq j'$
\beq
[\sigma_j(k),\sigma_{j'}(k')]=[\sigma_j(-k),\sigma_{j'}(k')]
=[\sigma_j(-k),\sigma_{j'}(-k')]=0 \ .
\eeq
All other commutators can be derived via
$\sigma_i^\dagger(-k)=\sigma_i^{\phantom\dagger}(k)$. 
The vacuum $|\Omega\rangle$ is defined by
\begin{equation}
\sigma_1(-k)|\Omega\rangle=\sigma_2(k)|\Omega\rangle=0
\label{def_vac}
\end{equation}
for all $k>0$. 
The notion of the {\em dual field} $\Theta(x)$ will also be useful. $\Theta(x)$ is defined by 
\begin{equation}
\partial_x \Theta(x)=-\Pi(x) \ ,
\label{dualfield}
\end{equation}
leading to the commutator
\begin{equation}
[\Theta(x),\phi(x')]=i\theta(x-x') \ .
\end{equation}
In terms of normal modes one finds
\begin{equation}
\Theta(x)=-\frac{i}{\sqrt{4\pi}} {\sum_{k\neq 0}} \,\frac{\sqrt{|k|}}{k}\, e^{-ikx}
\left(\sigma_1(k)-\sigma_2(k)\right) \ .
\end{equation}
The concept of {\em vertex operators} will play an important role in the sequel.
Vertex operators $V_j(\alpha;x)$ are defined as normal--ordered exponentials
\begin{equation}
V_j(\alpha;x)=\: :\exp\left(\pm\alpha \sum_{p\neq 0} \frac{\sqrt{|p|}}{p}\,
e^{-\frac{a}{2}|p|-ipx} \sigma_j(p) \right):
\label{vertexoperator}
\end{equation}
with $+$ (upper sign) corresponding to $j=1$ and $-$ (lower sign) to $j=2$. 
This sign convention will be used throughout this paper.
Normal ordering~$:\ldots :$ amounts to commuting all the operators that annihilate the vacuum
according to (\ref{def_vac}) to the right. One can 
rewrite (\ref{vertexoperator}) in terms of the field and its dual (\ref{dualfield}) 
\begin{equation}
V_j(\alpha;x)=\: :\exp\left(\pm i\alpha\sqrt{\pi} \int d\epsilon\,c(\epsilon)
\left[\phi(x+\epsilon)\pm\Theta(x+\epsilon)\right]\right): 
\end{equation}
with the Lorentzian
\begin{equation}
c(\epsilon)=\frac{a/2\pi}{\epsilon^2+a^2/4} \ .
\end{equation}
$c(\epsilon)$ is normalized 
\beq 
\int_{-\infty}^\infty  d\epsilon\:c(\epsilon)=1
\eeq 
and $c(\epsilon)\stackrel{a\rightarrow 0}{\longrightarrow} \delta(\epsilon)$. 
Further properties of vertex operators, in particular their operator 
product expansion (OPE), are reviewed in the Appendix.

One can rewrite the
interaction term of the sine--Gordon model (\ref{sinegordon1}) in terms of vertex operators
\begin{equation}
H=\int dx\, \left( \frac{1}{2}\Pi^2(x)+\frac{1}{2}\left(\frac{\partial\phi}{\partial x}\right)^2
+\frac{u}{2\pi a^2} \left(\frac{2\pi a}{L}\right)^{\alpha^2}
\Big(V_1(\alpha;x) V_2(-\alpha;x)+ V_2(\alpha;x) V_1(-\alpha;x) \Big) \right) \ .
\label{sinegordon2}
\end{equation}
The prefactor $(2\pi a/L)^{\alpha^2}$ follows from (\ref{rel_normalordering}).
Here and in the sequel $\alpha$ and $\beta$ are used interchangeably with the identification
\begin{equation}
\alpha\stackrel{\rm def}{=}\frac{\beta}{\sqrt{4\pi}} \ . 
\end{equation}
Eq.~(\ref{sinegordon2}) is the form of
the sine--Gordon Hamiltonian that we will investigate with the flow equation approach:
{\em No} implicit momentum cutoff is implied in (\ref{sinegordon2}): The UV--regularization 
of the Hamiltonian (\ref{sinegordon2}) is achieved by the cutoff parameter~$a>0$ in the 
vertex operators. 
The regularizations in (\ref{sinegordon2}) and (\ref{sinegordon1}) are related by
$a^{-1}\propto \Lambda \propto \tau^{-1}$. For a direct comparison between the original
Hamiltonian (\ref{sinegordon1}) and the form (\ref{sinegordon2}) used here one can also 
identically express (\ref{sinegordon2}) as
\begin{equation}
H=\int dx\, \left( \frac{1}{2}\Pi^2(x)+\frac{1}{2}\left(\frac{\partial\phi}{\partial x}\right)^2
+\frac{u}{\pi a^2} \cos\left[ \beta\int 
d\epsilon\,c(\epsilon)\,\phi(x+\epsilon) \right]\right) \ .
\label{sinegordon3}
\end{equation}
The regularization with $a$ therefore amounts to {\em smearing out} the interaction
term. In the limit $a\rightarrow 0$ one recovers the $\cos(\beta\phi(x))$--interaction term.

The universal properties of the sine--Gordon model for energies $|E|\ll a^{-1}$  
are not affected by this choice
of regularization. We find, however, notational 
simplifications and more compact expressions in the course of our calculation 
when we start with (\ref{sinegordon2}) (or equivalently (\ref{sinegordon3})). In order
to clarify the main conceptual ideas of the flow equation approach, it will therefore be
convenient for us to use the regularization (\ref{sinegordon2}) with the UV--cutoff~$a^{-1}$
built in via the definition of the vertex operators.

\subsection{Perturbative scaling analysis} \noindent
The flow equation approach can be viewed as an extension of perturbative scaling.
Therefore it is useful to briefly review the results of the perturbative scaling
analysis as applied to the sine--Gordon model. A comprehensive review can be
found in Ref.~\cite{Amit80}.

In 2-loop order there are two renormalization group equations that describe the flow 
of $u$ and $\beta$ upon integrating out the degrees of freedom with 
$\Lambda-d\Lambda < |k| <\Lambda$ (see Ref.~\cite{Wiegmann78})
\begin{eqnarray}
\frac{d\beta^{-2}}{d\ln\Lambda}&=&-\frac{u^2}{4\pi} \nonumber \\
\frac{du}{d\ln\Lambda}&=&\left(\frac{\beta^2}{4\pi}-2\right)\, u \ .
\label{pert_RG}
\end{eqnarray}
\begin{figure}[t]
\begin{center}
\leavevmode
\epsfxsize=10cm
\epsfysize=12cm
\epsfbox{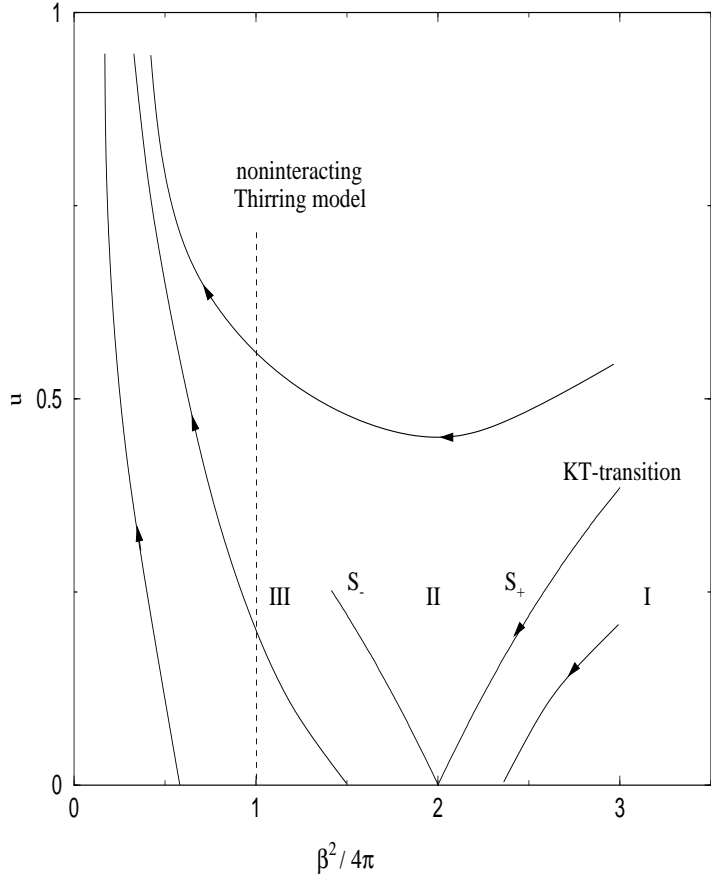}
~\vspace*{0.5cm}~

\caption{Perturbative scaling flow and the phase diagram of the sine--Gordon model.
The strong--coupling phase is to the left of the KT--transition line~$S_+$, the
weak--coupling phase to the right of~$S_+$.}
\label{fig_KT}
\end{center}
\end{figure}
The initial conditions are $u(\tau^{-1})=u$ and $\beta(\tau^{-1})=\beta$. These scaling equations
give rise to the Kosterlitz--Thouless type phase diagram shown in Fig.~\ref{fig_KT}.
The two separatrices~$S_{\pm}$ originating from $\beta^2=8\pi$ with $\beta^2=8\pi(1\pm u)$ for
small~$u$ divide the parameter space in three sectors:
\begin{enumerate}
\item the weak--coupling sector I;
\item the crossover sector II;
\item the asymptotic freedom sector III.
\end{enumerate}
Both in II and III the perturbative scaling equations (\ref{pert_RG}) lead to {\em strong--coupling}
behavior with the running coupling constant~$u$ growing larger and larger during the flow. 
Therefore the perturbative RG approach eventually becomes invalid in these sectors.
This indicates the opening of mass gap in the spectrum in II and III. Across~$S_+$ the
system undergoes a Kosterlitz--Thouless type phase transition between this massive phase 
and the massless phase in~I. 

In spite of the strong--coupling divergence in II and III, the perturbative scaling equations
allow us to analyze the size of the mass gap by identifying the mass~$M$ with the scaling
invariant of~(\ref{pert_RG}). One e.g.\ finds the following expressions for small $u>0$
\begin{eqnarray}
M\propto \Lambda\, \left( \frac{u}{2-\beta^2/4\pi} \right)^{1/(2-\beta^2/4\pi)}
&\qquad&\mbox{for~} \frac{1-\beta^2/8\pi}{u} \gg 1  
\label{scaling1} \\
M\propto \Lambda\, \exp\left( -\frac{1}{2-\beta^2/4\pi} \right)
&\qquad&\mbox{for~} \left( 1-\frac{\beta^2}{8\pi}-u \right)^{-1} \gg 1
\quad\mbox{(along~} S_-)
\label{2loop} \\
M\propto \Lambda\, \exp\left( -\frac{\pi}{2\sqrt{ u^2-(1-\beta^2/8\pi)^2 }} \right)
&\qquad&\mbox{for~} \left( 1-\frac{\beta^2}{8\pi}+u \right)^{-1} \gg 1
\quad\mbox{(along~} S_+)
\label{scaling2}
\end{eqnarray}
In this manner one can obtain information about the renormalized low--energy scale even in the
strong--coupling phase. But the perturbative scaling approach does by itself not lead to an
understanding of the physical behavior associated with this low--energy scale. This situation is
typical for other strong--coupling problems as well, the Kondo model being the paradigm
in condensed matter theory \cite{Wilson75}. In combination with mappings to other exactly
solvable models these shortcomings can sometimes be partially overcome, see Sect.~II.C below. 
However, in general there
is considerable interest in theoretical methods that can solve strong--coupling problems in a controlled
way. Therefore the flow equation approach might be an interesting tool also for other strong--coupling
problems by removing some of the above shortcomings. 

The phase diagram Fig.~\ref{fig_KT} remains essentially unchanged in higher loop orders~\cite{Amit80}. For
latter comparison with the flow equation solution it is interesting to also write down the
3-loop result for the mass gap on $S_-$ (that is for $\beta^2=8\pi(1-u)$) in the limit 
$u\rightarrow 0$
\begin{equation}
M\propto \Lambda\, u^{1/2} \exp\left(-\frac{1}{2u}\right) \ .
\label{3loop}
\end{equation}
Notice the $u^{1/2}$--prefactor as compared to the 2-loop result (\ref{2loop}). 
Higher loop orders beyond 3-loop should only affect the proportionality factor in (\ref{3loop})~\cite{Amit80}.

\subsection{Integrable structure and relation to other models} \noindent
The sine--Gordon model is one of the best studied integrable models, which
makes it a very suitable test model for our new approach. Its spectrum was obtained
exactly from an inverse scattering solution~\cite{Faddeev75} and its $S$--matrix
was calculated by Zamolodchikov~\cite{Zamolodchikov77}. For a recent review
see Ref.~\cite{Gogolin98}.

In the strong--coupling phase the exact solution confirms the scaling behavior
(\ref{scaling1}) for the mass~$M$ of the solitons and antisolitons. The exact
$S$--matrix~\cite{Zamolodchikov77} also shows that for small
rapidity differences these solitons and antisolitons behave as fermions. 
This important observation of a {\em change in statistics} for the low--energy
excitations will be reproduced in our flow equation framework. 

In addition, the exact solution shows that new features appear for
$\beta^2<4\pi$: Soliton--anitsoliton bound states ({\em breathers}) emerge
in the spectrum with excitation energies smaller than $2M$. There is one breather
for $8\pi/3\leq\beta^2<4\pi$, two breathers for 
$2\pi\leq\beta^2<8\pi/3$~etc.~\cite{Gogolin98}.

The sine--Gordon model is related to other integrable models like the
spin-1/2 X-Y-Z chain and the Thirring model. Since in particular the relation to
the Thirring model will be fruitful in the sequel, it is useful to sum
up some of its main properties here:
The massive $1d$ Thirring model is defined by the Lagrangian density
\begin{equation}
{\cal L}_{\rm Th}=\sum_{\mu=0}^1\left( \bar\psi i\gamma_\mu\partial^\mu \psi
- \frac{1}{2}\,g\, j_\mu j^\mu\right) -M\bar\psi \psi 
\label{thirring}
\end{equation}
in terms of the two--component spinor~$\psi(x)$
\begin{equation}
\psi(x)=\left( \begin{array}{r} \psi_1(x) \\ \psi_2(x) \end{array} \right) 
\end{equation}
with components obeying fermionic anticommutation relations
\beq
\{\psi^\ph_j(x),\psi^\dagger_{j'}(y)\}=\delta_{jj'}\delta(x-y) \ , \quad
\{\psi^\ph_j(x),\psi^\ph_{j'}(y)\}
=\{\psi^\dagger_j(x),\psi^\dagger_{j'}(y)\}=0 \ .
\eeq
The current is defined by $j^\mu\stackrel{\rm def}{=}\bar\psi\gamma^\mu \psi$ with
$\bar\psi \stackrel{\rm def}{=} \psi^\dagger \gamma_0$ and
the $\gamma$--matrices are explicitly given by
\begin{equation}
\gamma_0=\left( \begin{array}{rr} 0 & 1 \\ 1 & 0 \end{array} \right) \quad\mbox{and}\quad
\gamma_1=\left( \begin{array}{rr} 0 & 1 \\ -1 & 0 \end{array} \right) \ .
\end{equation}
The exact Bethe ansatz solution of the Thirring model was obtained by Bergknoff
and Thacker~\cite{Bergknoff79}.

Coleman~\cite{Coleman75} has shown the equivalence of the sine--Gordon model
with the Thirring model (\ref{thirring}) order by order in perturbation theory with the following
mapping between the coupling constants
\begin{equation}
\frac{\beta^2}{4\pi}=\frac{1}{1+g/\pi} \ .
\end{equation}
One notices that $\beta^2=4\pi$ is a special point of the sine--Gordon model since it corresponds
to a {\em noninteracting} massive Thirring model ($g=0$). For $\beta^2=4\pi$ the elementary
excitations of the sine--Gordon model are therefore {\em fermionic} with the dispersion relation
$\pm E_k$ with
\begin{equation}
E_k=\sqrt{k^2+M^2} \ .
\label{dispersion_Thirring}
\end{equation}
The explicit relation between the bosonic field of the sine--Gordon model and the fermionic
field of the Thirring model was found by Mandelstam~\cite{Mandelstam75}. For $\beta^2=4\pi$
one has explicitly 
\begin{equation}
\psi_j(x)=\frac{1}{\sqrt{L}} V_j(-1;x) 
\end{equation}
with $V_j(\alpha;x)$ from (\ref{vertexoperator}): Notice that the $V_j(\pm 1;x)$ obey 
anticommutation relations (\ref{vertex_fermion})
in the limit $a\rightarrow 0$. These {\em Thirring fermions}\footnote{Notice, however, 
that $V_1(\pm 1;x)$ commutes with 
$V_2(\pm 1;x)$ instead of anticommuting. ``Proper'' fermions can easily be defined 
with an additional Jordan--Wigner phase factor, but nothing new can be learned from this 
Jordan--Wigner construction.} correspond to the quantized soliton
solutions of the sine--Gordon model~\cite{Mandelstam75}.

The perturbative scaling approach does not ``know'' about the special point $\beta^2=4\pi$
where the sine--Gordon model is trivially diagonalizable by using the equivalence to the
quadratic Thirring model: The strong--coupling scaling trajectories in Fig.~\ref{fig_KT}  
go right through the line $\beta^2=4\pi$. The subsequent strong--coupling divergence of 
the running coupling constant
is then due to the fact that one has generated the nonvanishing energy scale~$M$. 
A similar scenario occurs in the Kondo model: There the diagonal
Hamiltonian corresponds to the Toulouse point \cite{Toulouse69} and the nonvanishing energy
scale is the Kondo temperature~$T_K$. 

A standard approach to avoid the strong--coupling divergence is to scale the model
to the exactly solvable line $\beta^2=4\pi$: One stops the scaling once 
$\beta^2(\Lambda_{\rm eff})=4\pi$ and obtains the mass 
from the value of the running coupling constant. With this approach it is also plausible 
that the low--energy single--particle/hole excitations in the strong--coupling
phase are fermionic with a mass set by the scaling invariant. Though very useful,
this is an uncontrolled approximation since the running coupling constant is
already large when $\beta^2(\Lambda_{\rm eff})=4\pi$. It is therefore difficult/impossible to
learn something about the crossover from weak--coupling to strong--coupling or about the
effect of irrelevant operators at the strong--coupling fixed point. These shortcomings 
make it desirable to develop our new method that will allow a systematic expansion
describing the full crossover flow.

The sine--Gordon model is also related to a variety of other models like
the $2d$~Coulomb gas with temperature $T=\beta^{-2}$ and fugacity $z\propto u$,
or a $1d$~electron gas with backward scattering. For an overview of these and other relations 
see Ref.~\cite{Solyom79}. As a final remark we also want to mention that
the mapping to the $1d$~electron gas gives a natural
interpretation to the separatrices~$S_\pm$ in Fig.~\ref{fig_KT} since they 
correspond to an electron gas with SU(2) spin--symmetric interactions~\cite{Solyom79}.
The sine--Gordon model with $\beta^2=8\pi(1\pm u)$, $|u|\ll 1$ therefore carries a
hidden SU(2)--symmetry. 

\section{Flow equation approach}
\subsection{General concepts} \noindent
The idea to apply a sequence of infinitesimal unitary transformations to a
Hamiltonian in order to make it more diagonal has been independently
put forward by Wegner~\cite{Wegner94} and G{\l}azek and Wilson~\cite{GlazekWilson}.
Wegner's original work focussed on diagonalizing many--particle
Hamiltonians, whereas the focus in the work of G{\l}azek and Wilson was to
construct effective low--energy Hamiltonians for strong--coupling field
theories: Such effective Hamiltonians can then be analyzed by standard
techniques in order to find the bound state spectrum, which in turn could be
interpreted as e.g.\ hadrons or mesons. Though the outlook of these
approaches is somehow different, the concepts are very similar.
In this paper we will follow Wegner's methodology.

The main idea of Wegner's {\em flow equations} is to generate a one--parameter
family of unitarily equivalent Hamiltonians $H(B)$ labelled by a {\em flow
parameter}~$B$.\footnote{The flow parameter has been denoted
by~$\ell$ in most other works on flow equations. In order to avoid
confusion with the common notation where $\ell$ is the logarithm of
the change in length scale in RG--equations, $B$ instead of $\ell$ is used in this work.}
This is achieved by solving a differential equation
\begin{equation}
\frac{dH(B)}{dB}=[\eta(B),H(B)]
\label{flow_etaH} 
\end{equation}
with some anti-Hermitian generator $\eta(B)=-\eta^\dagger(B)$ where 
$H(B=0)=H$ is the initial Hamiltonian. One wants to choose $\eta(B)$ 
such that $H(B)$ becomes more diagonal as $B\rightarrow\infty$: Splitting 
up $H(B)$ in its diagonal and interaction parts
\begin{equation}
H(B)=H_0(B)+H_{\rm int}(B) \ ,
\end{equation}
this amounts to requiring $H_{\rm int}(B)$ becomes (in some sense) smaller for 
$B\rightarrow\infty$. In order to achieve this, Wegner proposed the following
generator~\cite{Wegner94}
\begin{equation}
\eta(B)\stackrel{\rm def}{=}[H_0(B),H_{\rm int}(B)] \ .
\label{canonical_eta}
\end{equation}
With this choice of $\eta(B)$ one can show 
\begin{equation}
\frac{d}{dB}\,{\rm Tr}\,H_{\rm int}^2(B) \leq 0 
\label{trace}
\end{equation}
and in this sense the operator $H_{\rm int}(B)$ becomes smaller along the flow. Notice
that $B$ has the dimension of (Energy)${}^{-2}$ with this choice. However, for
a many--particle Hamiltonian Eq.~(\ref{trace}) is usually not well--defined
since the trace is typically infinite. Also 
higher and higher order interactions are successively generated by the system
of equations (\ref{flow_etaH}) and (\ref{canonical_eta}), which have to be truncated
in some way making rigorous statements difficult. 

Still one finds that (\ref{canonical_eta}) is generally a suitable choice for 
achieving our goal to make the initial Hamiltonian diagonal if $H_{\rm int}(B=0)$
can be viewed as a small perturbation term: Truncating the system of higher order
interactions produced by (\ref{flow_etaH}) and (\ref{canonical_eta}) in some order of 
the coupling constant then amounts to generating a perturbation expansion in
a renormalized coupling constant. From this point of view the flow equation 
approach is similar to perturbative RG. Matrix elements of $H_{\rm int}(B=0)$
that couple states with large energy differences are eliminated in the initial
stages of the flow (for small $B$), and matrix elements coupling more degenerate
states are eliminated later. This is reminiscent of the energy scale separation
underlying the renormalization group approach, which is the suitable
perturbation expansion for systems with largely varying energy scales.

Explicit applications of these ideas have been discussed for various model
Hamiltonians like the $n$--orbital model \cite{Wegner94,KabelWegner97},
dissipative quantum systems \cite{KehreinMielke97},  
systems with electron--phonon coupling \cite{Mielke97,RagwitzWegner99} 
and various other models in condensed matter 
physics \cite{GlazekWilson98,KehreinMielke96,KnetterUhrig00,Stein00,Stein99,CremersMielke99}.
In the present paper it will be shown how this method can be used
for the sine--Gordon model as
a genuine strong--coupling many--body Hamiltonian.

\subsection{Generator $\eta$} \noindent
The aim of this work is to diagonalize the sine--Gordon model (\ref{sinegordon2})
using the method of infinitesimal unitary transformations outlined above. We split
up the sine--Gordon Hamiltonian $H(B)$ into a free part $H_0$ and the interaction
part $H_{\rm int}(B)$ 
\begin{equation}
H(B)=H_0+H_{\rm int}(B)
\end{equation}
with
\begin{eqnarray}
H_0&=& \int dx\, \left( \frac{1}{2}\Pi^2(x)+\frac{1}{2}\left(\frac{\partial\phi}{\partial x}\right)^2 
\right) \nonumber \\
&=& \sum_{p>0} p\,\left(\sigma_1(p)\sigma_1(-p)+\sigma_2(-p)\sigma_2(p) \right) \\
H_{\rm int}(B)&=& \int dx\,dy\,u(B;y)\,\left(
V_1(\alpha;x) V_2(-\alpha;x-y) + {\rm h.c.} \right) \ .
\label{Hint}
\end{eqnarray}
In order to avoid confusion the initial parameters $u$ and $\beta$ in (\ref{sinegordon2})
will from now on be denoted as $u_0$ and $\beta_0$. The initial condition for (\ref{Hint}) 
then reads
\begin{eqnarray}
\alpha&=&\frac{\beta_0}{\sqrt{4\pi}} \nonumber \\
u(B=0;y)&=& \frac{u_0}{2\pi a^2}\, \delta(y) \left(\frac{2\pi a}{L}\right)^{\alpha^2} \ .
\label{initialuBx}
\end{eqnarray} 
Notice that we have already allowed for a general nonlocal interaction $u(B;y)$ in (\ref{Hint})
since the initially local interaction (\ref{initialuBx}) will become nonlocal along the 
flow (see below). Next we have to evaluate (\ref{canonical_eta}). The following commutator
is useful
\begin{equation}
[\sigma_j(p),V_{j'}(\alpha;x)]=\delta_{jj'}\,\alpha\,\frac{\sqrt{|p|}}{p}\,
\exp\left(-\frac{a}{2}|p|+ipx\right) \, V_{j'}(\alpha;x) 
\end{equation}
leading to
\begin{eqnarray}
{[\,\sum_{p>0}p\, \sigma_1(p)\sigma_1(-p),V_1(\alpha;x)]} &=& i\partial_x V_1(\alpha;x) \nonumber \\
{[\,\sum_{p>0}p\, \sigma_2(-p)\sigma_2(p),V_2(\alpha;x)]} &=& -i\partial_x V_2(\alpha;x) \ .
\label{comm2}
\end{eqnarray}
Thus we find the following generator
\begin{eqnarray}
\eta(B)&=&[H_0,H_{\rm int}(B)] \nonumber \\
&=&-2i\int dx\,dy\,\frac{\partial u(B;y)}{\partial y}\,\big(
V_1(\alpha;x) V_2(-\alpha;x-y) + {\rm h.c.} \big) \ . 
\label{def_eta}
\end{eqnarray}

\subsection{Commutator $[\eta,H_0]$} \noindent
To study the flow generated by $\eta$ we first look at the commutator $[\eta,H_0]$.
Using (\ref{comm2}) one easily shows
\begin{equation}
[\eta,H_0]=4 \int dx\,dy\,\frac{\partial^2 u(B;y)}{\partial y^2}\,\big(
V_1(\alpha;x) V_2(-\alpha;x-y) + {\rm h.c.} \big) \ . 
\label{etaH0}
\end{equation}
Comparison of the coefficients on the left--hand side of (\ref{flow_etaH}) with
(\ref{etaH0}) gives
\begin{equation}
\frac{\partial u(B;y)}{\partial B}=4\,\frac{\partial^2 u(B;y)}{\partial y^2} \ ,
\label{diff}
\end{equation}
where possible contributions from $[\eta,H_{\rm int}]$ are still missing. 
Eq.~(\ref{diff}) has the character of a diffusion equation:
The initially local interaction becomes increasingly non--local along the flow. In 
terms of Fourier coefficients
\begin{equation}
u(B;y)=\sum_p u(B;p) e^{-ipy} 
\label{FT_u}
\end{equation}
one finds the solution
\begin{equation}
u(B;p)=\frac{u_0}{4\pi^2 a^2}\, e^{-4p^2 B} \left(\frac{2\pi a}{L}\right)^{\alpha^2} \ .
\label{u_decay_1}
\end{equation}
One sees explicitly that matrix elements $u(B;p)$ coupling states with large energy
differences~$|p|$ are eliminated in the early stages of the flow (for small $B$), whereas
matrix elements coupling more degenerate states are decoupled later during the flow.
This is a generic feature of Wegner's generator (\ref{canonical_eta}). 

Later we will see that (\ref{u_decay_1}) is modified due to higher--order contributions. 
Therefore we introduce a more general parametrization 
\begin{equation}
u(B;p)=\frac{\tilde u(B)}{4\pi^2 a^2} \left(\frac{2\pi a}{L}\right)^{\alpha^2(B)} v(B;p) 
\label{u_decay_2}
\end{equation} 
with a {\em running coupling}~$\tilde u(B)$, initially $\tilde u(B=0)=u_0$. 
The differential equation for the coefficients $v(B;p)$ now reads
\begin{equation}
\frac{\partial v(B;p)}{\partial B}=-4p^2 v(B;p)
\label{free_v}
\end{equation}
with the initial condition $v(B=0;p)=1$.

\subsection{Commutator $[\eta,H_{\rm int}]$}
\subsubsection{General properties} \noindent
The evaluation of $[\eta,H_{\rm int}]$ is the key calculation in the flow equation approach.
We first look at some general properties of such commutators. Let $A_1, A_2, B_1, B_2$ be
arbitrary operators with $[A_j,B_{j'}]=0$. We define $*O*$ as the operator with its ground
state expectation value subtracted
\begin{equation}
*O* \stackrel{\rm def}{=} O- \langle O \rangle \ ,
\label{def_star}
\end{equation}
with the notation $\langle O \rangle\stackrel{\rm def}{=} \langle\Omega| O|\Omega \rangle$.
One easily shows
\begin{eqnarray}
{[A_1 B_1, A_2 B_2]}&=& 
\langle A_1 A_2\rangle \langle B_1 B_2\rangle - \langle A_2 A_1\rangle \langle B_2 B_1\rangle 
\label{A1B1A2B2} \\
&&+ \langle B_1 B_2\rangle *A_1 A_2* - \langle B_2 B_1\rangle *A_2 A_1* \nonumber \\
&&+ \langle A_1 A_2\rangle *B_1 B_2* - \langle A_2 A_1\rangle *B_2 B_1* \nn \\
&&+R  \nonumber
\end{eqnarray}
with 
\begin{equation}
R=  *A_1 A_2*\, *B_1 B_2* - *A_2 A_1*\, *B_2 B_1* \ .
\end{equation}
In general $R$ leads to the generation of higher order interaction terms during the flow. 
$R$ vanishes if the operators fulfill the following exchange relations
\begin{eqnarray}
A_1 A_2 +e^{i\phi} A_2 A_1 &=& c \nonumber \\
B_1 B_2 +e^{-i\phi} B_2 B_1 &=& c
\label{no_R} 
\end{eqnarray}
with fixed $\phi$ and $c$.
E.g.\ for $\phi=0$ these are fermionic anticommutation relations, or for $\phi=\pi$
bosonic commutation relations. Then no higher--order interactions are generated and it is
possible to close the flow equations without approximations. For general $\beta_0$ in the
interaction term we will, however, have to develop 
a suitable approximation for~$R$ in the next section.

\subsubsection{$[\eta,H_{\rm int}]$ in the sine--Gordon model} \noindent
There are two structurally different commutators of vertex operators generated
by $[\eta,H_{\rm int}]$ in the sine--Gordon model: $[V^\ph_1 V^\dagger_2, V^\dagger_1 V^\ph_2]$
and $[V^\ph_1 V^\dagger_2,V^\ph_1 V^\dagger_2]$ (or equivalently 
$[V^\dagger_1 V^\ph_2,V^\dagger_1 V^\ph_2]$). The first term 
\begin{equation}
[V_1(\alpha;x_1) V_2(-\alpha;x_1-y_1), V_1(-\alpha;x_2) V_2(\alpha;x_2-y_2)] 
\end{equation}
and its hermitian conjugate
will turn out to be the leading contributions and are discussed first. 
Eqs.~(\ref{vertex_expectationvalue1}) and (\ref{vertex_expectationvalue2})
give
\begin{eqnarray}
\langle V_1(\alpha;x_1) V_1(-\alpha;x_2)\rangle = s_1^{-\alpha^2} \ &,&\ 
\langle V_1(-\alpha;x_2) V_1(\alpha;x_1)\rangle =\bar s_1^{-\alpha^2} \nonumber \\
\langle V_2(-\alpha;x_1-y_1) V_2(\alpha;x_2-y_2)\rangle = s_2^{-\alpha^2} \ &,&\ 
\langle V_2(\alpha;x_2-y_2) V_2(-\alpha;x_1-y_1)\rangle = \bar s_2^{-\alpha^2} 
\end{eqnarray}
with 
\begin{eqnarray}
s_1&=& \frac{2\pi}{L} \left(i(x_2-x_1)+a\right) \nonumber \\
s_2&=& \frac{2\pi}{L} \left(i(x_1-y_1-x_2+y_2)+a\right) \ .
\label{def_s}
\end{eqnarray}
Using (\ref{A1B1A2B2}) we then find
\begin{eqnarray}
\lefteqn{ [V_1(\alpha;x_1) V_2(-\alpha;x_1-y_1), V_1(-\alpha;x_2) V_2(\alpha;x_2-y_2)] } 
\label{vertex_comm} \\
&=& s_1^{-\alpha^2} s_2^{-\alpha^2} - \bar s_1^{-\alpha^2} \bar s_2^{-\alpha^2} \nonumber \\
&& s_2^{-\alpha^2} *V_1(\alpha;x_1) V_1(-\alpha;x_2)*
-\bar s_2^{-\alpha^2} *V_1(-\alpha;x_2) V_1(\alpha;x_1)* \nonumber \\
&& s_1^{-\alpha^2} *V_2(-\alpha;x_1-y_1) V_2(\alpha;x_2-y_2)*
-\bar s_1^{-\alpha^2} *V_2(\alpha;x_2-y_2) V_2(-\alpha;x_1-y_1)* \nonumber \\
&&+R \nonumber
\end{eqnarray}
with
\begin{eqnarray}
R&=& *V_1(\alpha;x_1) V_1(-\alpha;x_2)*\,*V_2(-\alpha;x_1-y_1) V_2(\alpha;x_2-y_2)* \nonumber \\
&& -*V_1(-\alpha;x_2) V_1(\alpha;x_1)*\,*V_2(\alpha;x_2-y_2) V_2(-\alpha;x_1-y_1)* \ .
\end{eqnarray}
The key approximation in our method is to use an operator product expansion (OPE) in
higher--order interaction terms like~$R$, and then to neglect contributions with larger
scaling dimensions (more irrelevant terms in the RG--sense). From (\ref{vertex_OPE1}) we e.g.\ 
conclude 
\begin{equation}
*V_1(\alpha;x_1) V_1(-\alpha;x_2)* = s_1^{-\alpha^2} \left(i\alpha(x_2-x_1)
\sum_{p \neq 0} \sqrt{|p|} e^{-\frac{a}{2}|p|-ipx_1} \sigma_1(p) + \ldots \right) \ ,
\label{V1_OPE}
\end{equation}
where we have neglected the higher order terms in (\ref{vertex_OPE1}). Notice that the
c--number contribution has already been removed by subtracting the ground state expectation
value. Putting everything together gives
\begin{equation}
R=\alpha^2(x_2-x_1)(x_1-y_1-x_2+y_2)(s_1^{-\alpha^2} s_2^{-\alpha^2}
-\bar s_1^{-\alpha^2} \bar s_2^{-\alpha^2})
\sum_{p,q \neq 0} \sqrt{|p\,q|} e^{-\frac{a}{2}(|p|+|q|)-ipx_1-iq(x_1-y_1)} 
\sigma_1(p) \sigma_2(q) \ .
\end{equation}
The first and second term in (\ref{vertex_comm}) are c--numbers and describe a shift in
the ground state energy. This is of no particular interest and we will not look into it.  
The various other terms generated in $[\eta,H_{\rm int}]$ are discussed in the next subsections.

\subsubsection{$R$--term} \noindent
The $R$--term in (\ref{vertex_comm}) leads to the following contribution from
$[\eta,H_{\rm int}]$
\begin{eqnarray}
&&-2i\int dx_1 dx_2 dy_1 dy_2\,\frac{\partial u(B;y_1)}{\partial y_1} u(B;y_2) \nonumber \\
&&\times 2\alpha^2 (x_2-x_1)(x_1-y_1-x_2+y_2)(s_1^{-\alpha^2} s_2^{-\alpha^2}
-\bar s_1^{-\alpha^2} \bar s_2^{-\alpha^2})
\sum_{p,q \neq 0} \sqrt{|p\,q|} e^{-\frac{a}{2}(|p|+|q|)-ipx_1-iq(x_1-y_1)} 
\sigma_1(p) \sigma_2(q) \nonumber \\
&=&-8\pi i\alpha^2 \left(\frac{L}{2\pi}\right)^{2\alpha^2} \sum_{k \neq 0} |k| t_k\: \sigma_1(k) \sigma_2(-k) 
\end{eqnarray}
with coefficients $t_k$
\begin{equation}
t_k=\int dz_1 dz_2 dz_3 \,e^{-a|k|-ikz_1} \frac{\partial u(B;z_1)}{\partial z_1} u(B;z_1+z_2)
z_3(z_2-z_3)\left((iz_3+a)^{-\alpha^2}(i(z_2-z_3)+a)^{-\alpha^2} -{\rm h.c.}\right) \ .
\end{equation}
Except for an (unimportant) initial transient where $B\lesssim a^2$, the $z_1$--integral leads to
the following expression
\begin{equation}
t_k=-2\pi i a^{4-2\alpha^2} \sum_{p} p\, u(B;p) u(B;-k-p) \int dx\, e^{i(k+p)ax}
\:I(x)
\label{wk}
\end{equation}
with
\begin{equation}
I(x)=\int dy\,y(x-y)\left( (1+iy)^{-\alpha^2}(1-iy+ix)^{-\alpha^2}
- (1-iy)^{-\alpha^2}(1+iy-ix)^{-\alpha^2} \right) \ .
\end{equation}
Writing
$1+i\left(y\pm\frac{x}{2}\right)=r_\pm e^{i\phi_\pm}$
with 
\bea
r_\pm&=&\sqrt{1+\left(y\pm \frac{x}{2}\right)^2} \nn \\
\phi_\pm &=&\arcsin\frac{y\pm\frac{x}{2}}{r_\pm} \in [-\frac{\pi}{2},\frac{\pi}{2}] 
\eea
leads to
\beq 
I(x)=2i \int dy\,\left(\frac{x^2}{4}-y^2\right) (r_+ r_-)^{-\alpha^2}
\sin\left(\alpha^2 (\phi_--\phi_+)\right) \ .
\eeq 
The flow is dominated by the term decaying most slowly with~$B$, which corresponds
to the large-$x$ behavior of $I(x)$. Therefore we can approximate
\beq
r_\pm \approx \left|\frac{x}{2}\right|\,|z\pm 1| \ ,
\eeq
where $z=2y/x$, and find
\beq
I(x)=2i\left|\frac{x}{2}\right|^{3-2\alpha^2} \int_{-\infty}^\infty dz\:
(1-z^2)\,|1-z^2|^{-\alpha^2} \sin(\alpha^2(\phi_- -\phi_+)) \ .
\eeq
With the above approximation one has $\phi_\pm\in \{-\frac{\pi}{2},\frac{\pi}{2}\}$.
Thus the only contributions to $I(x)$ come from regions with $\phi_-\neq\phi_+$
\bea
\phi_- -\phi_+=-\pi &\quad\Rightarrow\quad & x>0,~ -1<z<1 \\
\phi_- -\phi_+=\pi &\quad\Rightarrow\quad & x<0,~ -1<z<1 \nn
\eea
leading to
\bea
I(x)&=&-2i\sin(\alpha^2\pi) {\rm sgn}(x) \left|\frac{x}{2}\right|^{3-2\alpha^2}
\int_{-1}^1 dz\,(1-z^2)^{1-\alpha^2} \nn \\
&=&2i\: \frac{\pi^{3/2}}{\Gamma(\alpha^2-1)\,\Gamma(\frac{5}{2}-\alpha^2)} \:
{\rm sgn}(x) \left|\frac{x}{2}\right|^{3-2\alpha^2} \ .
\eea
With (\ref{u_decay_2}) and (\ref{free_v}) the sum over $p$ in (\ref{wk}) gives
\beq
\sum_{p} p\, u(B;p) u(B;-k-p) \, e^{ipax}
=i\left(\frac{\tilde u(B)}{4\pi^2 a^2}\right)^2 \left(\frac{2\pi a}{L}\right)^{2\alpha^2}
\sqrt{\frac{\pi}{8B}} \left(\frac{ax}{16B}+\frac{ik}{2}\right)
\,\exp\left(-\frac{a^2 x^2}{32B}-\frac{ikax}{2}-2Bk^2\right) \ .
\eeq
The final step is to perform the $x$--integration in (\ref{wk}). This can be done
in closed form leading to hypergeometric functions. However, the flow is
determined by the IR--limit $k\rightarrow 0$ where the integral is simpler
\beq
t_{k=0}=i\,\frac{32\pi^3}{\Gamma(\alpha^2-1)} (32B)^{1-\alpha^2}
\left(\frac{\tilde u(B)}{4\pi^2 a^2}\right)^2 \left(\frac{2\pi a}{L}\right)^{2\alpha^2} \ .
\eeq
For the full $k$--dependence we write
\beq
t_k=t_{k=0}\,f(\alpha^2;k\sqrt{B}) \ .
\eeq
\begin{figure}[t]
\begin{center}
\leavevmode
\epsfxsize=8cm
\epsfysize=8cm
\epsfbox{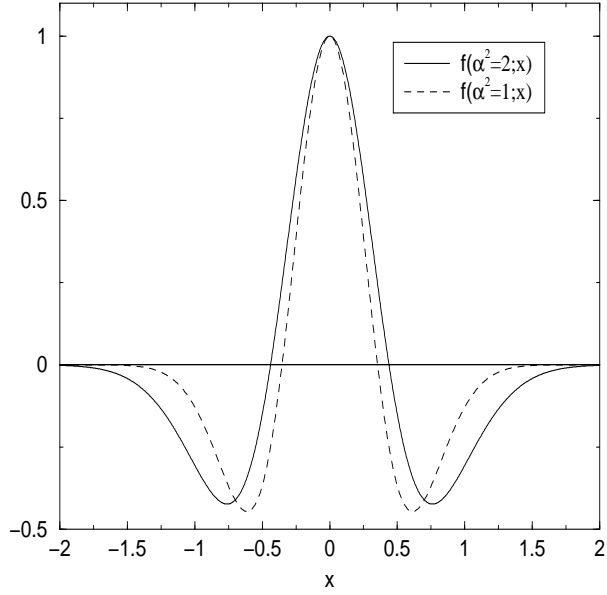}
~\vspace*{1cm}

\caption{$f(\alpha^2;x)$ for $\alpha^2=2$ and $\alpha^2=1$, see Eq.~(\protect\ref{def_f}).}
\label{fig_f}
\end{center}
\end{figure}
To leading order the only information that we will need about $f(\alpha^2;x)$
is that it falls of rapidly to zero for large arguments~$|x|\gg 1$. For example 
one easily shows
\bea
f(\alpha^2=1;x)&=&e^{-4x^2}(1-8x^2) \nn \\
f(\alpha^2=2;x)&=&e^{-4x^2}-\sqrt{2\pi}\: x\:
{\rm erf}(\sqrt{2}x)\,e^{-2x^2} \ .
\label{def_f}
\eea
These functions are depicted in Fig.~\ref{fig_f}.

Putting everything together the $R$--terms in (\ref{vertex_comm}) from
$[\eta,H_{\rm int}]$ contribute
\beq
[\eta,H_{\rm int}]\longrightarrow
\frac{32}{a^2}\left(\frac{32B}{a^2}\right)^{1-\alpha^2} \frac{\alpha^2}{2\Gamma(\alpha^2-1)}\, \tilde u^2(B)\: 
\sum_{k \neq 0} |k|\, f(\alpha^2;k\sqrt{B})\, \sigma_1(k) \sigma_2(-k) \ .
\label{sigma1sigma2}
\eeq
An important observation can be made for $\alpha=1$: Due to the divergent $\Gamma$--function in the denominator,
the term (\ref{sigma1sigma2}) vanishes for $\beta_0^2=4\pi$. This is an immediate consequence of the fact that 
in this case the vertex operators describe fermions. The interaction term of the sine--Gordon model is then simply 
a quadratic term in the fermions and no higher order interactions are generated during the flow, 
therefore according to (\ref{no_R}) $R\equiv 0$ for all~$B$. 
For $\beta_0^2=4\pi$ we will be able to solve the flow equations without any approximations, thereby recovering
the equivalence to the noninteracting Thirring model discussed in Sect.~II.C. This demonstrates a fundamental
difference of our approach to perturbative scaling, where the scaling trajectories go right through
the line $\beta^2=4\pi$ (see Fig.~\ref{fig_KT}).

\subsubsection{$H_{\rm diag}$} \noindent
Let us next look at the fifth and sixth term in (\ref{vertex_comm}). The total contribution from
$[\eta,H_{\rm int}]$ to terms of this structure is
\bea
[\eta,H_{\rm int}]&\longrightarrow&
-2i\int dx_1 dx_2 dy_1 dy_2\,\frac{\partial u(B;y_1)}{\partial y_1} u(B;y_2) 
\label{Hnew1} \\
&&\times\bigg( s_1^{-\alpha^2}\, *V_2(-\alpha;x_1-y_1) V_2(\alpha;x_2-y_2)*
-\bar s_1^{-\alpha^2}\, *V_2(\alpha;x_2-y_2) V_2(-\alpha;x_1-y_1)* \nn \\
&&\quad +s_1^{-\alpha^2}\, *V_2(\alpha;x_1-y_1) V_2(-\alpha;x_2-y_2)* 
-\bar s_1^{-\alpha^2}\, *V_2(-\alpha;x_2-y_2) V_2(\alpha;x_1-y_1)*\bigg) \nn \ .
\eea
For simplicity we will only look at the first term in this expression since all the other terms can be treated
likewise. We first exchange the two vertex operators using (\ref{reflection}) as this will lead to a 
normal--ordered expression below
\beq
-2i\int dx_1 dx_2 dy_1 dy_2\,\frac{\partial u(B;y_1)}{\partial y_1} u(B;y_2)  \:
s_1^{-\alpha^2}\,\frac{\bar s_2^{\alpha^2}}{s_2^{\alpha^2}}\: *V_2(\alpha;x_2-y_2) V_2(-\alpha;x_1-y_1)* 
\eeq
with $s_1, s_2$ from (\ref{def_s}). We can rewrite this in terms of Fourier transforms
($\alpha>0$)
\beq
V_j(-\alpha;x)\stackrel{\rm def}{=} \sum_p e^{ipx} V_j(-\alpha;p) \ , \quad 
V_j(\alpha;p)\stackrel{\rm def}{=} \left[V_j(-\alpha;p)\right]^\dagger \ ,
\label{fourier} 
\eeq
substitute $x_2\rightarrow x_2+x_1$ and perform the integral over~$x_1$. This leads to
\bea
&&-4\pi i\left(\frac{L}{2\pi}\right)^{\alpha^2}\int dx_2 dy_1 dy_2 \,\frac{\partial u(B;y_1)}{\partial y_1} u(B;y_2)  \nn \\ 
&&\times\sum_k  *V_2(\alpha;k) V_2(-\alpha;k)* \:
(ix_2+a)^{-\alpha^2} \frac{[i(x_2-y_2+y_1)+a]^{\alpha^2}}{[-i(x_2-y_2+y_1)+a]^{\alpha^2}} \:
e^{-ik(x_2+y_1-y_2)} \nn \\
&=& \ldots \nn \\
&=&-8\pi^2 \left(\frac{L}{2\pi}\right)^{\alpha^2} 
\int dx\, dy \sum_{k,p} p\,u^2(B;p)\,e^{-ikx+ipy}  *V_2(\alpha;k) V_2(-\alpha;k)* \:
[i(x+y)+a]^{-\alpha^2} \frac{[ix+a]^{\alpha^2}}{[-ix+a]^{\alpha^2}} \ ,
\eea
where we have employed (\ref{FT_u}).
Next the $y$--integration can be done using
\beq
\int dy\,(iy+ix+a)^{-\alpha^2} e^{ipy} = e^{-ipx} |p|^{\alpha^2-1} \frac{2\pi}{\Gamma(\alpha^2)}\,\theta(p) \ ,
\eeq
which is valid
in the limit $|ap|\ll 1$: This holds except for an (unimportant) initial transient with wavevectors
of order $a^{-1}$. We find
\beq
-\frac{16\pi^3}{\Gamma(\alpha^2)} \left(\frac{L}{2\pi}\right)^{\alpha^2} 
\int dx\, \sum_{k} \sum_{p>0} p^{\alpha^2}\,u^2(B;p)\,e^{-i(k+p)x}  *V_2(\alpha;k) V_2(-\alpha;k)* \:
\frac{[ix+a]^{\alpha^2}}{[-ix+a]^{\alpha^2}} \ .
\eeq
Next we can approximate 
\beq
\frac{[ix+a]^{\alpha^2}}{[-ix+a]^{\alpha^2}} \rightarrow \cos(\pi\alpha^2)+i\sin(\pi\alpha^2)\,{\rm sgn}(x) 
\eeq
using the same reasoning as in (\ref{approx_vertex}). This gives
\bea
-\frac{16\pi^3}{\Gamma(\alpha^2)} \left(\frac{L}{2\pi}\right)^{\alpha^2} &&
\bigg( 2\pi\cos(\pi\alpha^2) \sum_{k>0} k^{\alpha^2}\,u^2(B;k)\: *V_2(\alpha;-k) V_2(-\alpha;-k)* \nn \\
&&+i\sin(\pi\alpha^2) \sum_{k} \sum_{p>0} p^{\alpha^2}\,u^2(B;p)\int dx\,e^{-i(k+p)x} {\rm sgn}(x)
*V_2(\alpha;-k) V_2(-\alpha;-k)* \bigg)
\nn \\
=-2a^{\alpha^2-4}\frac{\tilde u^2(B)}{\Gamma(\alpha^2)} 
\left(\frac{2\pi a}{L}\right)^{\alpha^2} &&
\bigg( \cos(\pi\alpha^2) \sum_{k>0} k^{\alpha^2}\,v^2(B;k)\: *V_2(\alpha;-k) V_2(-\alpha;-k)* \nn \\
&&+\frac{1}{\pi}\sin(\pi\alpha^2) \sum_{k} 
\sum_{\begin{array}{l} \scriptstyle p>0 \\ \scriptstyle p\neq k \end{array} } 
p^{\alpha^2}\,v^2(B;p) \, \frac{1}{p-k} \:
*V_2(\alpha;-k) V_2(-\alpha;-k)* \bigg)
\label{step1}
\eea
In order to do the sum over~$p$ in the second term it will be sufficient to use the approximate
solution $v(B;p)=e^{-4Bp^2}$ from (\ref{free_v}). Deviations from this approximate solution 
essentially only occur close to the strong--coupling fixed point $\alpha^2=1$, where the 
second term vanishes anyway. This $p$--summation leads to an integral of the type
\bea
h(\alpha^2;x)&=&
{\rm P}\int_0^\infty dy\: e^{-y^2}\,\frac{y^{\alpha^2}}{y-x} \nn \\
&=&\frac{1}{2}\,e^{-x^2} |x|^{\alpha^2} {\rm Re}\left\{ i^{\alpha^2}\,\left[
\Gamma\left(1+\frac{\alpha^2}{2}\right)\,\Gamma\left(-\frac{\alpha^2}{2},-x^2\right)
-i\,{\rm sgn}(x)\,\Gamma\left(\frac{1+\alpha^2}{2}\right)\,
\Gamma\left(\frac{1-\alpha^2}{2},-x^2\right) \right] \right\} \ ,
\label{def_h}
\eea
where $\Gamma(s,z)$ denotes the incomplete $\Gamma$--function. One easily shows
$h(\alpha^2;x=0)=\Gamma(\alpha^2/2)/2$ and the asymptotic behavior for $|x|\gg 1$
\beq
h(\alpha^2;x)=-\frac{\Gamma\left(\frac{1+\alpha^2}{2}\right)}{2x}\: +O(x^{-2}) 
\eeq
with a smooth crossover in between. 

It will be convenient to use the normalized operators
\bea
S_{j}(\alpha;k)&\stackrel{\rm def}{=}& \left[ \frac{2\pi}{L}\,\Gamma(\alpha^2) 
\left(\frac{L|k|}{2\pi}\right)^{1-\alpha^2} \right]^{1/2}
V_j(-\alpha;k) 
\label{def_Sk} \\
\Rightarrow\quad S^\dagger_{j}(\alpha;k)&=&\left[ \frac{2\pi}{L}\,\Gamma(\alpha^2) 
\left(\frac{L|k|}{2\pi}\right)^{1-\alpha^2} \right]^{1/2}
V_j(\alpha;k)  \nn
\eea
with the properties (see (\ref{prop_S}))
\beq
S_1^\dagger(\alpha;-k)|\Omega\rangle=S_1^\ph(\alpha;k)|\Omega\rangle
=S_2^\dagger(\alpha;k)|\Omega\rangle=
S_2^\ph(\alpha;-k)|\Omega\rangle=0 \quad \forall k>0 
\label{annihilation}
\eeq
and the normalization 
\bea
\langle S_1^\ph(\alpha;k) S_1^\dagger(\alpha;k') \rangle 
=\langle S_2^\dagger(\alpha;k) S_2^\ph(\alpha;k') \rangle
&=& \delta_{kk'} \theta(k)\, L/2\pi \nn \\
\langle S_1^\dagger(\alpha;k) S_1^\ph(\alpha;k') \rangle 
=\langle S_2^\ph(\alpha;k) S_2^\dagger(\alpha;k') \rangle
&=& \delta_{kk'} \theta(-k)\, L/2\pi  
\label{normalization}
\eea
for $|ak|, |ak'|\ll 1$
as follows easily from (\ref{vertex_expectationvalue1}) and
(\ref{vertex_expectationvalue2}).
We express (\ref{step1}) in terms of these operators and find
\bea 
- \,\frac{2\tilde u^2(B)}{a^3\, \Gamma^2(\alpha^2)}&\times& \bigg(
\cos(\pi\alpha^2) \sum_{k>0} \, (ak)^{2\alpha^2-1} v^2(B;k) \:
S_2^\dagger(\alpha;-k) S_2^\ph(\alpha;-k) \nn \\
&&+\frac{1}{\pi}\sin(\pi\alpha^2) \sum_{k>0} (ak)^{\alpha^2-1}
(8B/a^2)^{-\alpha^2/2} h(\alpha^2;\sqrt{8B}k)\:
S_2^\dagger(\alpha;-k) S_2^\ph(\alpha;-k) \nn \\
&&+\frac{1}{\pi}\sin(\pi\alpha^2) \sum_{k>0} (ak)^{\alpha^2-1}
(8B/a^2)^{-\alpha^2/2} h(\alpha^2;-\sqrt{8B}k)\:
*S_2^\dagger(\alpha;k) S_2^\ph(\alpha;k)*\bigg)
\label{S2}
\eea
In the first two terms we do not need the subtraction operation $*~*$ anymore since the 
vacuum is already annihilated by them. The third term does not yet annihilate the
vacuum. This can be easily achieved by using (\ref{vertex_exchange}). However,
already the second term will turn out to have hardly any effect, and the third term
is again smaller than the second term. In order to simplify our expressions we 
therefore omit the third term in the sequel, although there would be no problem at all
in carrying it along as well. 
Let us now also collect the other terms from (\ref{Hnew1}) leading to
\bea
[\eta,H_{\rm int}]&\longrightarrow&
- \,\frac{4\tilde u^2(B)}{a^3\, \Gamma^2(\alpha^2)} \sum_{k>0} \left(
\cos(\pi\alpha^2) \, (ak)^{2\alpha^2-1} v^2(B;k)
+\frac{1}{\pi}\sin(\pi\alpha^2) (ak)^{\alpha^2-1}
(8B/a^2)^{-\alpha^2/2} h(\alpha^2;\sqrt{8B}k) \right) \nn \\
&&\qquad\qquad\times \left( S_1^\ph(\alpha;-k) S_1^\dagger(\alpha;-k)
+ S_1^\dagger(\alpha;k) S_1^\ph(\alpha;k)
+ S_2^\dagger(\alpha;-k) S_2^\ph(\alpha;-k)+ S_2^\ph(\alpha;k) S_2^\dagger(\alpha;k) \right) \ .
\label{step2}
\eea
Since $\alpha$ generically flows as a function of~$B$, this implies that vertex
operators with different scaling dimensions contribute to each wavevector~$k$.
However, most of the contribution to a given $k$--vector occurs in a narrow
range of the flow: We will see in Sect.~IV.C that for a given~$k$ the main
contribution occurs when $B\approx B_k$ with 
\bea
B_k&\stackrel{\rm def}{=}&\frac{1}{4k^2} \qquad\qquad\qquad \mbox{(weak--coupling phase)} \nn \\
B_k&\stackrel{\rm def}{=}&\frac{1}{4k\sqrt{k^2+M^2}} 
\qquad \mbox{(strong--coupling phase with mass gap~$M$)} \ .
\label{def_Bk}
\eea
In order to simplify our notation we therefore use a single
scaling dimension~$\alpha$ corresponding to each~$k$ with $\alpha=\alpha(B_k)$.\footnote{
This approximation becomes exact in the low--energy limit, e.g.\ in the strong--coupling
phase for $|k|\ll M$.} The newly generated term in the Hamiltonian can then
be written as
\beq
H_{\rm diag}(B)=\sum_{k>0} \omega(B;k) \left( P_1^\ph(-k) P_1^\dagger(-k)+ P_1^\dagger(k) P_1^\ph(k)
+ P_2^\dagger(-k) P_2^\ph(-k)+ P_2^\ph(k) P_2^\dagger(k) \right)
\label{Hnew}
\eeq
with 
\beq
P_j^\ph(k)\stackrel{\rm def}{=} S_j^\ph(\alpha(B_k);k) \ , 
\quad
P_j^\dagger(k) = S_j^\dagger(\alpha(B_k);k)
\label{Sjk}
\eeq
and according to (\ref{step2})
\beq
\frac{\partial\omega(B;k)}{\partial B}
=- \,\frac{4\tilde u^2(B)}{a^3\, \Gamma^2(\alpha^2)}  \left(
\cos(\pi\alpha^2) \, (ak)^{2\alpha^2-1} v^2(B;k)
+\frac{1}{\pi}\sin(\pi\alpha^2) (ak)^{\alpha^2-1}
(8B/a^2)^{-\alpha^2/2} h(\alpha^2;\sqrt{8B}k) \right)
\eeq 
with $\omega(B=0;k)=0$.
Using Eqs.~(\ref{comm2}) one can easily check $[H_0,H_{\rm diag}(B)]=0$,
therefore $H_{\rm diag}(B)$ can be interpreted as diagonal. 

\subsubsection{$H_{\rm res}$} \noindent
So far we have not discussed the commutators with the structure
$[V^\ph_1 V^\dagger_2,V^\ph_1 V^\dagger_2]$ and $[V^\dagger_1 V^\ph_2,V^\dagger_1 V^\ph_2]$
that are also generated by $[\eta,H_{\rm int}]$. From (\ref{A1B1A2B2}) one concludes that these 
commutators contain only the $R$--term. The operator product expansion in the $R$--term
then generates interactions with the structure $V_1(2\alpha) V_2(-2\alpha)$ etc., that is
only terms with larger scaling dimensions. These interactions are neglected in the present
approximation, just like the higher--order terms in (\ref{V1_OPE}). We formally sum up
these neglected terms with larger scaling dimensions in $H_{\rm res}$. 

\subsection{Commutator $[\eta,H_{\rm diag}]$} \noindent
We also have to study the effect of the infinitesimal unitary transformations on the newly
generated terms (\ref{Hnew}). An overlap exists essentially only for wavevectors of 
order~$B^{-1/2}$. For notational simplicity we can therefore use the running scaling dimension
$\alpha=\alpha(B)$ in (\ref{Hnew}) and arrive at the follwing commutator
\bea
[\eta,H_{\rm diag}]&=& -2i \int dx\,dy\,\frac{\partial u(B;y)}{\partial y}\,\sum_{k>0} \omega(B;k) \\
&&\times\bigg( [V_1(\alpha;x)V_2(-\alpha;x-y), S_1^\ph(\alpha;-k) S_1^\dagger(\alpha;-k)+ S_1^\dagger(\alpha;k) S_1^\ph(\alpha;k) \nn \\
&&\qquad\qquad\qquad\qquad\qquad
+ S_2^\dagger(\alpha;-k) S_2^\ph(\alpha;-k)+ S_2^\ph(\alpha;k) S_2^\dagger(\alpha;k)]
+{\rm h.c.} \bigg) \ . \nn
\eea
A typical contribution comes e.g.\ from 
\beq
[V_1(\alpha;x),S_1^\dagger(\alpha;k) S_1^\ph(\alpha;k)] 
=V_1(\alpha;x) S_1^\dagger(\alpha;k) S_1^\ph(\alpha;k)
-S_1^\dagger(\alpha;k) *S_1^\ph(\alpha;k) V_1(\alpha;x)* 
-S_1^\dagger(\alpha;k) \langle S_1^\ph(\alpha;k) V_1(\alpha;x)\rangle \ .
\eeq
The first and the second term on the rhs lead to normal--ordered interactions with larger scaling 
dimensions and are therefore neglected (or formally contained in $H_{\rm res}$). The third term
on the rhs gives rise to interactions of the type $H_{\rm int}$ leading to
\bea
[\eta,H_{\rm diag}]&\longrightarrow&
-2i \int dx\,dy\,\frac{\partial u(B;y)}{\partial y}\,\sum_{k>0} \omega(B;k) \nn \\
&&\times\bigg( V_1(\alpha;x) S_2^\ph(\alpha;-k) \langle V_2(-\alpha;x-y) S_2^\dagger(\alpha;-k) \rangle 
-V_1(\alpha;x) S_2^\ph(\alpha;k) \langle S_2^\dagger(\alpha;k) V_2(-\alpha;x-y) \rangle \nn \\
&&\qquad + \langle V_1(\alpha;x) S_1^\ph(\alpha;-k) \rangle  S_1^\dagger(\alpha;-k) V_2(-\alpha;x-y)
- \langle S_1^\ph(\alpha;k) V_1(\alpha;x) \rangle  S_1^\dagger(\alpha;k) V_2(-\alpha;x-y) \nn \\
&&\qquad +{\rm h.c.} \bigg) \nn \\
&=&-2i \int dx\,dy\,\frac{\partial u(B;y)}{\partial y}\,\sum_{k>0} \omega(B;k)
\left[ \Gamma(\alpha^2) \frac{2\pi}{L} \left(\frac{L|k|}{2\pi}\right)^{1-\alpha^2} \right]^{-1/2} \nn \\
&&\times\bigg( V_1(\alpha;x) S_2^\ph(\alpha;-k) e^{-ik(x-y)}-V_1(\alpha;x) S_2^\ph(\alpha;k) e^{ik(x-y)} \nn \\
&&\qquad+ S_1^\dagger(\alpha;-k) V_2(-\alpha;x-y) e^{ikx} - S_1^\dagger(\alpha;k) V_2(-\alpha;x-y) e^{-ikx}  
+{\rm h.c.} \bigg) \nn \\
&=&-4\pi \int dx \sum_{k>0} \omega(B;k) \,k\,u(B;k)
\left[ \Gamma(\alpha^2) \frac{2\pi}{L} \left(\frac{L|k|}{2\pi}\right)^{1-\alpha^2} \right]^{-1/2} \nn \\
&&\times\bigg( V_1(\alpha;x) S_2^\ph(\alpha;-k) e^{-ikx}+V_1(\alpha;x) S_2^\ph(\alpha;k) e^{ikx} \nn \\
&&\quad
+ S_1^\dagger(\alpha;-k) V_2(-\alpha;x) e^{ikx} + S_1^\dagger(\alpha;k) V_2(-\alpha;x) e^{-ikx}   
+{\rm h.c.} \bigg) \nn \\
&=&-8\pi L \sum_{k} \omega(B;|k|) \,|k|\,u(B;k)
\frac{1}{\Gamma(\alpha^2)} \left(\frac{L|k|}{2\pi}\right)^{\alpha^2-1} 
\:\bigg( S_1^\dagger(\alpha;k) S_2^\ph(\alpha;k)+ S_2^\dagger(\alpha;k) S_1^\ph(\alpha;k) \bigg) 
\ .
\label{add_v}
\eea
This term has to be compared with (\ref{Hint})
\bea
H_{\rm int}&=& \int dx\,dy\,u(B;y)\,\left(
V_1(\alpha;x) V_2(-\alpha;x-y) + {\rm h.c.} \right) \nn \\
&=&2\pi L \sum_k u(B;k)  \frac{1}{\Gamma(\alpha^2)} \left(\frac{L|k|}{2\pi}\right)^{\alpha^2-1}
\bigg( S_1^\dagger(\alpha;k) S_2^\ph(\alpha;k)+ S_2^\dagger(\alpha;k) S_1^\ph(\alpha;k) \bigg) \ .
\eea 
Eq.~(\ref{add_v}) therefore gives an additional contribution to the previous flow equation
(\ref{free_v}) and one finds
\beq
\frac{\partial v(B;k)}{\partial B}
=-4k^2 v(B;k) -4|k|\, \omega(B;|k|) v(B;k) \ .
\eeq

\subsection{Flow of $\beta^2$} 
\subsubsection{Unitary transformation $e^U$} \noindent
We have seen that the term (\ref{sigma1sigma2})
\beq
\sum_k w_k(B) |k| \sigma_1(k) \sigma_2(-k) \:dB 
\label{newterm}
\eeq
with
\beq
w_k(B)=\frac{32}{a^2}\left(\frac{32B}{a^2}\right)^{1-\alpha^2(B)} \frac{\alpha^2(B)}{2\Gamma(\alpha^2(B)-1)}\, 
\tilde u^2(B)\: f(\alpha^2(B);k\sqrt{B})
\label{def_wk}
\eeq
is generated during the flow. This term is not contained in the original sine--Gordon 
Hamiltonian. We will now show how this term can be eliminated by an additional 
unitary transformation $e^U$ with
\begin{equation}
U=\sum_{p>0} \,\psi_p \, \left( \sigma_1(p) \sigma_2(-p) -\sigma_1(-p) \sigma_2(p) \right) 
\label{unitary_trf_U}
\end{equation}
with suitable parameters $\psi_p$~\cite{LutherEmery74}.

Let us first write down some general properties of this unitary transformation
for general $\psi_p$. The bosonic fields are transformed according to
\bea
e^{-U} \sigma_1(p) e^U &=& \sigma_1(p) \cosh\psi_p+\sigma_2(p) \sinh\psi_p \nn \\
e^{-U} \sigma_2(p) e^U &=& \sigma_2(p) \cosh\psi_p+\sigma_1(p) \sinh\psi_p
\label{trf_fields}
\eea
and vertex operators as
\bea
e^{-U} V_1(\alpha;x) e^U &=&e^C :\exp\left[ \alpha \sum_{p\neq 0} \cosh\psi_p 
\frac{\sqrt{|p|}}{p}\, e^{-\frac{a}{2}|p|-ipx} \sigma_1(p) \right] :\; 
:\exp\left[ \alpha \sum_{p\neq 0} \sinh\psi_p 
\frac{\sqrt{|p|}}{p}\, e^{-\frac{a}{2}|p|-ipx} \sigma_2(p) \right] : 
\label{trf_vertexoperators} \\
e^{-U} V_2(\alpha;x) e^U &=&e^C :\exp\left[- \alpha \sum_{p\neq 0} \cosh\psi_p 
\frac{\sqrt{|p|}}{p}\, e^{-\frac{a}{2}|p|-ipx} \sigma_2(p) \right] :\; 
:\exp\left[- \alpha \sum_{p\neq 0} \sinh\psi_p 
\frac{\sqrt{|p|}}{p}\, e^{-\frac{a}{2}|p|-ipx} \sigma_1(p) \right] : \ , \nn
\eea
where 
\beq
C=-\alpha^2\sum_{p>0} \sinh\psi_p\,\sinh\psi_{-p}\,\frac{e^{-a|p|}}{p} \ .
\label{defC}
\eeq

\subsubsection{$e^{-U} H e^U$} \noindent
We take the point of view that (\ref{newterm}) has been generated infinitesimally 
by integrating the flow equations from $B$ to $B+dB$.
We therefore apply the above unitary transformation (\ref{unitary_trf_U}) to $H$
\beq
H(B+dB) \longrightarrow e^{-U} H(B+dB) e^{U} \ .
\label{def_trfU}
\eeq
Let us first investigate the effect on $H_0$: For the choice
\beq
\psi_p=-\frac{w_p(B)}{2} dB
\eeq 
the transformation $e^{-U}H_0 e^U$ reproduces $H_0$ and generates an additional
term that annihilates (\ref{newterm}). This can be shown easily by using the
transformation rules (\ref{trf_fields}). Notice that terms of order
$\psi_p^2$ and higher can be neglected since $\psi_p$ is of order $dB$.

Next we have to find the effect of this transformation on $H_{\rm int}(B)$.
Using (\ref{trf_vertexoperators}) one finds 
\bea
e^{-U} V_1(\alpha;x) V_2(-\alpha;y) e^U
&=&V_1(\alpha;x) :\exp\left[ \alpha \sum_{p\neq 0} \psi_p 
\frac{\sqrt{|p|}}{p}\, e^{-\frac{a}{2}|p|-ipy} \sigma_1(p) \right]: \nn \\
&&\times V_2(-\alpha;y) :\exp\left[ \alpha \sum_{p\neq 0} \psi_p 
\frac{\sqrt{|p|}}{p}\, e^{-\frac{a}{2}|p|-ipx} \sigma_2(p) \right]: \ ,
\eea
where we have again used that $\psi_p$ is of order $dB$. Using an OPE,
we can combine the first two terms into a vertex operator with a modified
scaling dimension, and likewise for the second two terms. The calculation 
proceeds along similar lines as in (\ref{vertex_OPE1}):
\bea
\lefteqn{V_1(\alpha;x) :\exp\left[ \alpha \sum_{p\neq 0} \psi_p 
\frac{\sqrt{|p|}}{p}\, e^{-\frac{a}{2}|p|-ipy} \sigma_1(p) \right]: } \nn \\
&=&\exp\left[ \alpha \sum_{p>0} 
\frac{\sqrt{|p|}}{p}\, e^{-\frac{a}{2}|p|-ipx} \sigma_1(p) \right] \;
\exp\left[ \alpha \sum_{p<0} 
\frac{\sqrt{|p|}}{p}\, e^{-\frac{a}{2}|p|-ipx} \sigma_1(p) \right] \nn \\
&&\times \exp\left[ \alpha \sum_{p>0} \psi_p
\frac{\sqrt{|p|}}{p}\, e^{-\frac{a}{2}|p|-ipy} \sigma_1(p) \right] \;
\exp\left[ \alpha \sum_{p<0} \psi_p
\frac{\sqrt{|p|}}{p}\, e^{-\frac{a}{2}|p|-ipy} \sigma_1(p) \right] \nn \\
&=&e^C\times\exp\left[ \alpha \sum_{p>0} 
\frac{\sqrt{|p|}}{p}\, e^{-\frac{a}{2}|p|-ipx} (1+\psi_p e^{-ip(y-x)})
\sigma_1(p) \right] \nn \\
&&\times \exp\left[ \alpha \sum_{p<0} 
\frac{\sqrt{|p|}}{p}\, e^{-\frac{a}{2}|p|-ipx} (1+\psi_p e^{-ip(y-x)})
\sigma_1(p) \right]   
\label{flow_scale1}
\eea
where
\beq
C=-\alpha^2 \sum_{p>0} \psi_p\: \frac{\exp(-ap-ip(x-y))}{p} \ .
\eeq
From (\ref{def_f}) we know that $\psi_p$ falls of rapidly on an
energy scale of order~$B^{-1/2}$. The leading behavior of the sum
is therefore (except for an uninteresting initial transient)
\beq
C=-\alpha^2\psi_{p=0} \sum_{p>0} \frac{\exp(-\tilde a p-ip(x-y))}{p} 
\eeq
leading to (compare Eq.~(\ref{cnumber}))
\beq
e^C=\left(\frac{2\pi(\tilde a+i(x-y))}{L}\right)^{\alpha^2\psi_{p=0}} \ .
\label{flow_scale3}
\eeq
Here $\tilde a=s\sqrt{B}$ takes into account the UV--cutoff generated
by the decay of the functions $f(\alpha^2;x)$ from Fig.~\ref{fig_f}. The actual value
of the proportionality constant~$s$ will only affect our results in 
next to leading order as will be shown later. Still we give its value here
for latter comparison with the three loop scaling results: For
$\alpha^2=2$ one finds in a somehow lengthy calculation
\beq
\ln s=\ln2 +\frac{\pi}{4} -\frac{\gamma}{2} \ ,
\label{def_ln_s}
\eeq
where $\gamma\approx 0.577$ is Euler's constant.
$s$ depends only weakly on $\alpha^2$, therefore we will use this
value throughout. 
Next 
\bea
\lefteqn{\exp\left[ \alpha \sum_{p>0} 
\frac{\sqrt{|p|}}{p}\, e^{-\frac{a}{2}|p|-ipx} (1+\psi_p e^{-ip(y-x)})
\sigma_1(p) \right] } \nn \\
&=& \exp\left[ \alpha(1+\psi_{p=0}) \sum_{p>0} 
\frac{\sqrt{|p|}}{p}\, e^{-\frac{a}{2}|p|-ipx}
\sigma_1(p) \right] \nn \\
&&\times  \exp\left[ i(x-y)\alpha\psi_{p=0} \sum_{p>0} 
\sqrt{|p|}\, e^{-\frac{a}{2}|p|-ipx}
\sigma_1(p) +O((x-y)^2)\right] \nn \\
&=&V_1(\alpha(1+\psi_{p=0});x)\;
\times\left(1+i(x-y)\alpha\psi_{p=0}\sum_{p>0} \sqrt{|p|}\, e^{-\frac{a}{2}|p|-ipx}
\sigma_1(p) +O((x-y)^2)\right) \nn \\
&=&V_1(\alpha(1+\psi_{p=0});x) +\mbox{more irrelevant terms}
\label{flow_scale2}
\eea
up to less singular (more irrelevant) terms. We formally include these 
more irrelevant terms (they have the structure
of products of vertex operators multiplied by derivatives of the bosonic
field) into $H_{\rm res}$
and neglect them from now on. It should also be noted that we have
approximated the flow of the scaling dimension in the vertex operator by
restricting ourselves to the flow in the IR--limit 
$\alpha\rightarrow \alpha(1+\psi_{p=0})$. At first sight this seems to be a problematic 
approximation since the functions $f(\alpha^2;x)$ in (\ref{def_f}) are nontrivial. 
However, the effect of the unitary transformations is {\em only}
cumulative on the low--energy scale where $f(\alpha^2;x)\stackrel{x\rightarrow 0}{=}1$.
Hence it is possible to restrict ourselves to the IR--limit.

Putting everything together we find
\bea
H_{\rm int}(B+dB) &\longrightarrow& e^{-U} H_{\rm int}(B+dB) e^U 
\label{flow_scale_4} \\
&=& \left(\frac{2\pi s\sqrt{B}}{L}\right)^{-\alpha^2(B) w_{p=0}(B) dB} \int dx\,dy\,
u(B;y) \nn \\
&&\qquad\qquad\times
\left[V_1\left(\alpha(1-\frac{w_{p=0}(B)}{2}\,dB);x\right) 
V_2\left(-\alpha(1-\frac{w_{p=0}(B)}{2}\,dB;x-y\right) + {\rm h.c.} \right] \nn 
\eea
Similar to the OPE above we have set $x-y=0$ in (\ref{flow_scale3}), that
is we have neglected less singular terms.

Finally, we have to investigate the effect of the additional
unitary transformation (\ref{def_trfU}) on $H_{\rm diag}$ from
(\ref{Hnew}). This is similar to the action on the interaction
term, except that here we find terms of the structure
\bea
e^{-U} V_1(\alpha;x) V_1(-\alpha;y) e^U
&=&V_1(\alpha;x) V_1(-\alpha;y)
:\exp\left[ \alpha \sum_{p\neq 0} \psi_p 
\frac{\sqrt{|p|}}{p}\, e^{-\frac{a}{2}|p|-ipx} \sigma_2(p) \right]: \nn \\
&&\qquad\times\: :\exp\left[- \alpha \sum_{p\neq 0} \psi_p 
\frac{\sqrt{|p|}}{p}\, e^{-\frac{a}{2}|p|-ipy} \sigma_2(p) \right]: 
\eea
and likewise for $j=2$. The third and fourth term can be combined
using an OPE (\ref{vertex_OPE2}) and one easily notices that
the only surviving term is a constant~1 since all other terms are
of order~$\psi_p^2=O(dB^2)$. Hence
\beq
H_{\rm diag}(B+dB) \longrightarrow e^{-U} H_{\rm diag}(B+dB) e^U = 
H_{\rm diag}(B+dB) \ .
\eeq
Summing up, we have looked at another infinitesimal unitary
transformation (\ref{def_trfU}) that acts on the Hamiltonian
during the flow equation procedure in addition to the generator
(\ref{def_eta}). An alternative viewpoint is to say that the full
generator of the flow now takes the structure
\bea
\eta_{\rm new}(B)&=&-2i\int dx\,dy\,\frac{\partial u(B;y)}{\partial y}\,\big(
V_1(\alpha;x) V_2(-\alpha;x-y) + {\rm h.c.} \big) \nn \\
&&+\frac{1}{2}\sum_p w_p(B)\,
\left(\sigma_1(p) \sigma_2(-p) -\sigma_1(-p) \sigma_2(p) \right) \ .
\label{def_fulleta}
\eea

\subsubsection{Flow equation for $\beta^2$} \noindent
From (\ref{flow_scale_4}) we can read of that the additional infinitesimal unitary
transformation generates a flow in the scaling dimension of the vertex operators
\bea
\alpha(B) &\longrightarrow& \alpha(B) \left(1-\frac{w_{p=0}(B)}{2}\,dB\right) \nn \\
\Rightarrow\qquad \frac{d\alpha^2}{dB}&=&-w_{p=0}(B) \,\alpha^2(B)
\label{flow_scale_5} \\
&=&-\frac{32}{a^2}\left(\frac{32B}{a^2}\right)^{1-\alpha^2(B)} 
\frac{\alpha^4(B)}{2\Gamma(\alpha^2(B)-1)}\, 
\tilde u^2(B) \ . \nn
\eea
We can already see that $\alpha^2=1$ is a fixed point of the flow equation
approach due to the diverging $\Gamma$--function in the denominator of
(\ref{flow_scale_5}). {\em This will be one of the key results of our new
approach.}

According to (\ref{flow_scale_4}) this flow of the scaling dimension now
{\em induces} a flow of the coupling constant $u(B;y)$
\bea
u(B;y) &\longrightarrow& u(B;y) 
\left(\frac{2\pi s\sqrt{B}}{L}\right)^{-\alpha^2(B) w_{p=0}(B)\,dB} \nn \\
&=& u(B;y) \left(1-\alpha^2(B) w_{p=0}(B)\,dB \,
\ln\left(\frac{2\pi s\sqrt{B}}{L}\right)\right) \nn \\
\Rightarrow\qquad \frac{du(B;y)}{dB}&=& u(B;y)\;
\frac{d\alpha^2}{dB} \; \ln\left(\frac{2\pi s\sqrt{B}}{L}\right) \ .
\eea
The solution is straightforward
\bea
u(B;y)&=&u(B=0;y)\;\exp\left(\int_0^B dB'\:
\frac{d\alpha^2}{dB'} \; \ln\left(\frac{2\pi s\sqrt{B'}}{L}\right) \right) \nn \\
&=&u(B=0;y)\;\exp\left(\int_0^B dB'\: \frac{d\alpha^2}{dB'} \; 
\ln\left(\frac{2\pi a}{L}\,\sqrt{\frac{32B'}{a^2}}\,\frac{s}{\sqrt{32}}\right) 
\right) \nn \\
&=&u(B=0;y)\:\left(\frac{2\pi a}{L}\right)^{\alpha^2(B)-\alpha^2(0)}
\:\left(\frac{s}{\sqrt{32}}\right)^{\alpha^2(B)-\alpha^2(0)} \:
\exp\left(\frac{1}{2}\int_0^B dB'\: \frac{d\alpha^2}{dB'}\:
\ln\left(\frac{32B'}{a^2}\right)\right) \ .
\eea
Using the parametrization (\ref{u_decay_2}) 
\begin{equation}
u(B;p)=\frac{\tilde u(B)}{4\pi^2 a^2} \left(\frac{2\pi a}{L}\right)^{\alpha^2(B)} v(B;p)
\label{flow_scale_6}
\end{equation} 
we see that this can be most conveniently expressed as a flow equation
for the {\em running coupling constant} $\tilde u(B)$ in (\ref{flow_scale_6})
\beq
\tilde u(B)=u_0\; \left(\frac{s}{\sqrt{32}}\right)^{\alpha^2(B)-\alpha^2(0)} \:
\exp\left(\frac{1}{2}\int_0^B dB'\: \frac{d\alpha^2}{dB'}\:
\ln\left(\frac{32B'}{a^2}\right)\right) \ .
\eeq
Introducing the dimensionless logarithmic flow parameter $\ell$
\beq
\ell\stackrel{\rm def}{=}\frac{1}{2} \ln\left(\frac{32 B}{a^2}\right) \ ,
\label{dimless_ell}
\eeq
one can show by partial integration
\beq
\frac{1}{2}\int_0^B dB'\: \frac{d\alpha^2}{dB'}\:
\ln\left(\frac{32B'}{a^2}\right)
=-\int_0^\ell d\ell'\:\alpha^2(\ell')
+\alpha^2(\ell)\,\ell \ .
\eeq
Using this we can sum up the results of this section in the following two equations:
\bea
\tilde u(\ell)&=&
u_0\; \left(\frac{s}{\sqrt{32}}\right)^{\alpha^2(\ell)-\alpha^2(0)} \:
\exp\left(-\int_0^\ell d\ell'\:\alpha^2(\ell')
+\alpha^2(\ell)\,\ell \right)
\label{flow_tilde_u} \\
\frac{d\alpha^2}{d\ell}&=&
-u_0^2\,\left(\frac{s^2}{32}\right)^{\alpha^2(\ell)-\alpha^2(0)}
\frac{\alpha^4(\ell)}{\Gamma(\alpha^2(\ell)-1)}\:
\exp\left(4\ell-2\int_0^\ell d\ell'\:\alpha^2(\ell')\right) \ .
\label{flow_alpha2}
\eea
{\em These two equations constitute the key results of this work.} Eq.~(\ref{flow_alpha2})
describes the flow of the scaling dimension under the flow equation procedure, and
from Eq.~(\ref{flow_tilde_u}) it follows how this {\em induces} the flow of the
{\em running coupling constant} $\tilde u(\ell)$. Therefore these equations will
serve as a generalization of the scaling equations derived in perturbative 
renormalization theory in Sect.~II.B. The value of the constant~$s$ in these equations
will turn out to affect our results only in next to leading order.

\section{Solution of the flow equations}
\subsection{Summary of the flow equations} \noindent
In this section we will sum up the results for the flow of the sine--Gordon
Hamiltonian under the effect of the infinitesimal unitary transformation
(\ref{def_fulleta}) as derived above. For general~$B$ the sequence $H(B)$ of 
unitarily equivalent Hamiltonians takes the form
\beq
H(B)=H_0+H_{\rm int}(B)+H_{\rm diag}(B)+H_{\rm res}(B) \ .
\label{full_H}
\eeq
Here
\bea
H_0&=&\int dx\, \left( \frac{1}{2}\Pi^2(x)+\frac{1}{2}\left(\frac{\partial\phi}{\partial x}\right)^2 \right) \\
H_{\rm int}(B)&=&\int dx\,dy\,u(B;y)\,\left(
V_1(\alpha(B);x) V_2(-\alpha(B);x-y) + {\rm h.c.} \right) 
\label{def_Hint2} \\
H_{\rm diag}(B)&=&\sum_{k>0} \omega(B;k) \left( P_1^\ph(-k) P_1^\dagger(-k)
+P_1^\dagger(k) P_1^\ph(k)
+ P_2^\dagger(-k) P_2^\ph(-k)+ P_2^\ph(k) P_2^\dagger(k) \right) \ ,
\eea
with $P_j(k)$ given by (\ref{Sjk})
\beq
P_j(k)= \left[ \frac{\Gamma(\alpha^2(B_k))}{2\pi L}
\left(\frac{L|k|}{2\pi}\right)^{1-\alpha^2(B_k)} \right]^{1/2}
\int dx\:e^{-ikx} V_j(-\alpha(B_k);x)
\eeq
and $B_k$ from (\ref{def_Bk}). The operators $P^\ph_j(k), P^\dagger_j(k)$ will turn 
out to be the soliton creation and annihilation operators. They
are normalized according to (\ref{normalization}). Notice 
$H_{\rm diag}(B)|\Omega\rangle=0$. 

$H_{\rm res}$ contains the neglected terms and will from now on be omitted in our analysis.
At any given $B$--scale these neglected terms have a larger 
scaling dimension than the interaction term $H_{\rm int}(B)$: They
are more irrelevant by at least two spatial derivatives. Notice that
$H_{\rm res}$ vanishes for $\beta^2=4\pi$ since then no approximations
are made.

Summing up the differential flow equations for the parameters in $H(B)$
we have
\bea
\frac{\partial v(B;k)}{\partial B}
&=&-4k^2 v(B;k) -4k\, \omega(B;k) v(B;k)
\label{flow_v} \\
\frac{\partial(a\,\omega(B;k))}{\partial (B/a^2)}
&=& - \,\frac{4\tilde u^2(B)}{\Gamma^2(\alpha^2)}  \left(
\cos(\pi\alpha^2) \, (ak)^{2\alpha^2-1} v^2(B;k)
+\frac{1}{\pi}\sin(\pi\alpha^2) (ak)^{\alpha^2-1}
(8B/a^2)^{-\alpha^2/2} h(\alpha^2;\sqrt{8B}k) \right) 
\label{flow_omega}
\eea
for $k>0$, and $v(B;-k)=v(B;k)$ symmetric in~$k$. 
For notational convenience we have written $\alpha(B)$ without 
its argument~$B$ in these equations. 
$h(\alpha^2;x)$ has been defined in (\ref{def_h}) and
\beq
u(B;y)=\frac{\tilde u(B)}{4\pi^2 a^2}
\left(\frac{2\pi a}{L}\right)^{\alpha^2(B)}
\sum_p v(B;p)\,e^{-ipy} \ .
\eeq
The initial conditions are $H_{\rm res}(B=0)=0$, $\omega(B=0;k)=0$ and 
$v(B=0;k)=1$, whereupon (\ref{full_H}) takes 
the form (\ref{sinegordon2}) of our original sine--Gordon Hamiltonian. 
For $\alpha^2_0<1$ one should in fact be more cautious and start the integration
only at $B=a^2$: The above equations hold only for $B\gtrsim a^2$ since we
have used $|ak|\gg 1$ in our calculation. One can easily verify that there is
no flow of the parameters for $B\ll a^2$, and the simplest way to take this
into account is to pose the initial conditions at $B=a^2$.

As we will see later the flow of the parameters is such that
$v(B=\infty;k)=0$ for $\beta_0^2>2\pi$. In this parameter region
the Hamiltonian $H(B)$ from (\ref{full_H}) therefore
becomes diagonal with $H_{\rm int}(B=\infty)=0$ 
in the limit $B\rightarrow\infty$ as expected under
the flow equation procedure
\beq
H(B=\infty)=H_0+H_{\rm diag}(B=\infty) \ ,
\eeq
notice $[H_0,H_{\rm diag}(B)]=0$.

The equations (\ref{flow_v}) and (\ref{flow_omega}) have to supplemented 
with the differential equation governing the flow of the scaling dimension 
(\ref{flow_alpha2})
and the thereby induced flow of the running coupling constant
(\ref{flow_tilde_u}).
It is possible to rewrite (\ref{flow_alpha2}) in a 
more conventional form for comparison with the perturbative 
scaling approach. Introducing a new function
\beq
u(\ell)\stackrel{\rm def}{=}u_0 
\left(\frac{s}{\sqrt{32}}\right)^{\alpha^2(\ell)-\alpha^2(0)}
\exp\left(2\ell-\int_0^\ell d\ell'\: \alpha^2(\ell') \right)
\eeq
we can rewrite (\ref{flow_alpha2}) as a set of two coupled 
differential equations
\bea
\frac{d\alpha^2}{d\ell}&=&-\frac{\alpha^4(\ell)}{\Gamma(\alpha^2(\ell)-1)}
\:u^2(\ell) 
\label{flow_alpha_fulla} \\
\frac{du}{d\ell}&=& \big(2-\alpha^2(\ell)\big)\: u(\ell)
+\left(\frac{\pi}{4}-\frac{\gamma}{2}-\frac{1}{2}\ln 8\right)\:
\frac{d\alpha^2}{d\ell} \: u(\ell) 
\label{flow_alpha_fullb}
\eea
with the initial conditions $u(\ell=0)=u_0$ and
$\alpha(\ell=0)=\beta_0/\sqrt{4\pi}$. For convenience we have used 
the dimensionless flow parameter $\ell=\frac{1}{2} \ln (32 B/a^2)$
from~(\ref{dimless_ell}) in these equations. Notice that the
second term in (\ref{flow_alpha_fullb}) is of order~$u^3$ and does 
therefore not contribute to the leading behavior for small~$u_0$. Since our 
present flow equation expansion has not taken all the terms in order~$u^3$
into account anyway, we can omit this term and arrive at
\bea
\frac{d\alpha^2}{d\ell}&=&-\frac{\alpha^4(\ell)}{\Gamma(\alpha^2(\ell)-1)}
\:u^2(\ell) \nn \\
\frac{du}{d\ell}&=& \big(2-\alpha^2(\ell)\big)\: u(\ell) \ .
\eea
The running coupling 
constant $\tilde u(\ell)$ from (\ref{flow_tilde_u}) can also be expressed as
\beq
\tilde u(\ell)=u(\ell)\:\exp\big( (\alpha^2(\ell)-2)\,\ell \big) \ .
\label{tilde_u1}
\eeq
In terms of the sine--Gordon parameter $\beta$ these equations take
the equivalent form 
\bea
\frac{d\beta^{-2}(\ell)}{d\ell}&=&\frac{1}{4\pi\Gamma\big(
-1+\beta^2(\ell)/4\pi\big)}\:u^2(\ell) \nn \\
\frac{du}{d\ell}&=&\left(2-\frac{\beta^2(\ell)}{4\pi}\right)\:u(\ell) 
\label{flow_beta}
\eea
and
\beq
\tilde u(\ell)=u(\ell)\:\exp\left( 
\left(\frac{\beta^2(\ell)}{4\pi}-2\right)\:\ell \right) \ .
\label{rel_u_tildeu}
\eeq
Finally it is of some interest to express $H_{\rm int}(B)$ directly in terms of 
the bosonic field $\phi(x)$ and its dual $\Theta(x)$ (see Eq.~(\ref{dualfield})).
After a short calculation one finds
\bea
H_{\rm int}(B)&=&\frac{\tilde u(B)}{\pi a^2} \int dx\,dy\,v(B;y)\:
\cos\left(\beta(B)\int d\epsilon\,c(\epsilon)\:
\frac{1}{2}\big(\phi(x+\epsilon)+\phi(x-y+\epsilon)
+\Theta(x+\epsilon)-\Theta(x-y+\epsilon)\big)\right) \ .
\eea
Since $v(B;y)$ becomes more and more nonlocal during the flow, one sees
that the interaction term of the sine--Gordon model evolves from the
original $\cos(\beta\phi(x))$--structure to a nonlocal interaction term 
with the structure
\beq
\cos\big((\beta/2)(\phi(x)+\phi(x-y)+\Theta(x)-\Theta(x-y))\big) \ .
\eeq
It is also possible to express $H_{\rm diag}(B)$ in terms of
$\phi(x)$ and $\Theta(x)$, however, this expression does not lead
to new insights. 

\subsection{Strong--coupling phase} 
\subsubsection{Fixed points and phase structure} 
\begin{figure}[t]
\begin{center}
\leavevmode
\epsfxsize=10cm
\epsfysize=10cm
\epsfbox{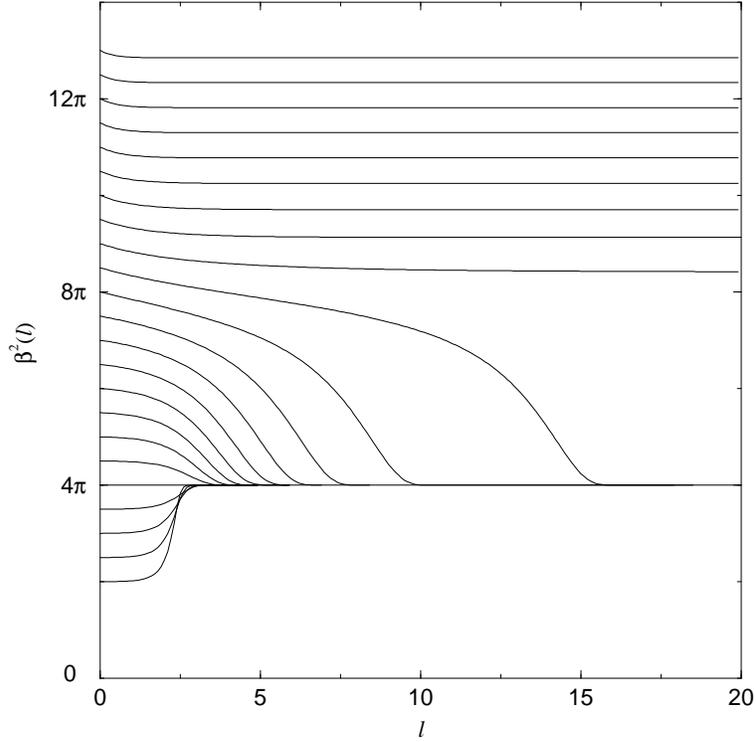}

\caption{Flow of $\beta^2(\ell)$ under the flow equation procedure
for $u_0=0.1$. Observe the strong--coupling fixed point $\beta^2(\ell=\infty)=4\pi$
and the weak--coupling fixed points with $\beta^2(\ell=\infty)\geq 8\pi$.}
\label{fig_flowbeta}
\end{center}
\end{figure}

\noindent
We will now work on the explicit solution of the flow equations
(\ref{flow_v}), (\ref{flow_omega}) and (\ref{flow_beta}). 
First we focus on the solution of (\ref{flow_beta}), since knowledge of the 
flow of $\beta(B)$ and $\tilde u(B)$ is necessary for solving the system
of equations (\ref{flow_v}) and (\ref{flow_omega}) later on.\footnote{Implicitly
an assumption about the behavior of $v(B;k)$ has been made when deriving 
(\protect\ref{flow_beta}) in Sect.~III.D. One can check that this assumption 
is self--consistently justified by analyzing the whole system of equations.}
From (\ref{flow_beta}) one concludes that there are two possible kinds of
asymptotic behavior: Either $\beta^2(\infty)=4\pi$ or $\beta^2(\infty)\geq 8\pi$.
$\beta^2(\infty)=4\pi$ will turn out to be the {\em attractive strong--coupling
fixed point} and values $\beta^2(\infty)\geq 8\pi$ correspond to the gapless
{\em weak--coupling phase}. These flows are depicted in Fig.~\ref{fig_flowbeta}.
Notice that the fundamental difference from the
perturbative scaling equations (\ref{pert_RG}) is the $\Gamma$--function in
the denominator of our flow equation for $\beta^{-2}(\ell)$. Therefore 
$\beta^2=4\pi$ is a fixed point in our approach which will be the main
difference as compared to perturbative RG. This does not come as
a surprise since the flow of $\beta^2$ followed from higher order terms
in the commutator $[\eta(B),H_{\rm int}(B)]$ in Sect.~III.D. However, for
$\beta^2=4\pi$ our interaction term is quadratic if considered as an
interaction term for Thirring fermions (see Sect.~II.C), and naturally
no higher order terms can be generated in $[\eta(B),H_{\rm int}(B)]$.
Another way of saying this is that the flow in $\beta^2$ is due to
approximations in the flow equation scheme when higher order terms
are generated. No flow of $\beta^2$ can occur if the flow equations close exactly.

In both phases $\tilde u(\ell)$ remains finite for $\ell\rightarrow\infty$.
On the other hand, one easily checks that $u(\ell)$ diverges in the 
strong--coupling phase and vanishes asymptotically in the weak--coupling
phase. Since $u(\ell)$ has so far only been introduced for rewriting 
(\ref{flow_alpha2}), its divergence in the strong--coupling phase need not
worry us here. The question of the true expansion parameter of
our approach will be discussed below in Sect.~IV.D, and we will see that this
expansion parameter is {\em not} $u(\ell)$. 

In order to analyze the phase boundaries we can expand the $\Gamma$--function
in (\ref{flow_beta}) around $\beta^2=8\pi$. In leading order this reproduces
the perturbative scaling equations (\ref{pert_RG})
\begin{eqnarray}
\frac{d\beta^{-2}}{d\ell}&=&\frac{u^2}{4\pi} \nonumber \\
\frac{du}{d\ell}&=&\left(2-\frac{\beta^2}{4\pi}\right)\, u \ .
\label{pert_RG2}
\end{eqnarray}
This approximation eventually breaks down in the strong--coupling
phase as $\beta^2(\ell)$ flows to $4\pi$: Then (\ref{pert_RG2}) is not a good
approximation for the true flow equation (\ref{flow_beta}) anymore.
Notice the sign difference from (\ref{pert_RG}) because $\ell$ from 
(\ref{dimless_ell}) corresponds to $-\ln\Lambda$, therefore $\ell$ is integrated
from 0 to $\infty$. Our flow equation approach therefore reproduces the
conventional two--loop scaling equations if we expand around $\beta^2=8\pi$.
In this way we also reproduce the hidden SU(2)--symmetry of
the sine--Gordon model for $\beta_0^2=8\pi(1\pm u_0)$ mentioned in
Sect.~II.C, although our approximation scheme does not manifestly 
respect this symmetry. Besides
showing the consistency of our new approach with the conventional perturbative 
RG scheme at $\beta^2\approx 8\pi$, we
also see immediately that our flow equations reproduce the
Kosterlitz--Thouless 
phase diagram Fig.~\ref{fig_KT} of the sine--Gordon model established with RG: 
In the limit of small initial~$u_0$ and $\beta_0^2>8\pi(1+u_0+O(u_0^2))$ we flow
to a weak--coupling fixed point, for $\beta_0^2<8\pi(1+u_0+O(u_0^2))$ to the 
strong--coupling fixed point $\beta^2(\infty)=4\pi$. These two phases
are again separated by a Kosterlitz--Thouless type transition 
along $\beta_0^2=8\pi(1+u_0+O(u_0^2))$. For the rest of this section we will 
focus on the strong--coupling phase. We will return to the weak--coupling 
phase in Sect.~IV.C.

\subsubsection{Low--energy effective Hamiltonian} \noindent
One advantage of the flow equation scheme is that we can easily analyze
the behavior of our model with the final diagonal Hamiltonian $H(B=\infty)$.
However, a simpler kind of analysis is possible by identifying a
low--energy effective Hamiltonian and analyzing this effective Hamiltonian. This will be done in
this subsection. Our results will be confirmed by the analysis of $H(B=\infty)$
in Sect.~IV.C later on, but the identification of a low--energy effective
Hamiltonian allows us to make contact with the conventional scaling picture,
which is very useful for providing a simple coherent description of the flow
equation approach in the strong--coupling phase. 

Let us look at $H(B)$ for large $B$ such that $|\beta^2(B)-4\pi| \ll 1$.
Then we can approximately set $\alpha(B)=1$ in $H_{\rm int}(B)$ from (\ref{def_Hint2}) 
and rewrite (\ref{full_H})
\beq
H(B)=H_{\rm eff}(B)+H_{\rm diag}(B) \ ,
\eeq
where
\beq
H_{\rm eff}(B)=\int dx\, \left( \frac{1}{2}\Pi^2(x)+\frac{1}{2}\left(
\frac{\partial\phi}{\partial x}\right)^2 \right)
+\frac{\tilde u(B)}{a} \sum_k v(B;k)\:
\big( P^\dagger_1(k) P^\ph_2(k) + P^\dagger_2(k) P^\ph_1(k) \big ) \ . 
\label{Heff}
\eeq
Now for $\alpha=1$ the creation and annihilation operators $P^\ph_j(k)$, $P^\dagger_j(k)$
from (\ref{Sjk}) obey fermionic anticommutation relations (\ref{vertex_fermion})
\beq
\{P_j^\dagger(k),P_j(k')\}=\delta_{k k'} \frac{L}{2\pi} \quad , \quad
\{P_j^\dagger(k),P_j^\dagger(k')\}=\{P_j(k),P_j(k')\}=0 \ .
\eeq
One could also rewrite the kinetic term $H_0$ in terms of these fermions
and would then arrive at a noninteracting Thirring model (\ref{thirring}) with a nonlocal 
mass term as the low--energy effective Hamiltonian in the strong--coupling phase. 
However, we can also analyze the spectrum of our low--energy effective Hamiltonian
$H_{\rm eff}(B)$ directly by working out the following commutators
\bea
{[H_{\rm eff}(B),P^\dagger_1(k)]} &=& k\,P^\dagger_1(k) 
+\frac{\tilde u(B)}{a} v(B;k) \,P^\dagger_2(k) \nn \\
{[H_{\rm eff}(B),P^\dagger_2(k)]} &=& -k\,P^\dagger_2(k) 
+\frac{\tilde u(B)}{a} v(B;k) \,P^\dagger_1(k)
\eea
leading to the dispersion relation 
\beq
E_k=\sqrt{k^2+\left(\frac{\tilde u(B)}{a}\,v(B;k)\right)^2} \ .
\eeq
This dispersion relation describes the single--particle/hole excitation spectrum
of the full Hamiltonian $H(B)$ for momenta $|k|\ll 1/\sqrt{B}$: According to
(\ref{flow_omega}) the terms in $H_{\rm diag}(B)$ corresponding to such momenta 
are only generated for even larger~$B$, therefore we can neglect the effect of 
$H_{\rm diag}(B)$ for excitations with $|k|\ll 1/\sqrt{B}$. In this limit we
also find the initial value $v(B;k)=1$ unchanged according to (\ref{flow_v}). 
Summing up, in the limit $k\rightarrow 0$
the dispersion relation for single--particle/hole
excitations in the strong--coupling phase has the form~$\pm E_k$ with
\beq
E_k=\sqrt{k^2+M^2}
\label{dispersion1}
\eeq
and the mass
\beq
M=\frac{\tilde u(\ell=\infty)}{a} \ .
\label{mass}
\eeq
Eq.~(\ref{dispersion1}) is also the form expected from exact methods using
integrability~\cite{Johnson73}. 
Equation (\ref{mass}) is a key result in this work since it describes
the relation between the running coupling and the generated mass
term in the strong--coupling regime.

We observe that the finiteness of the running coupling $\tilde u(\ell)$
in (\ref{Heff}) 
in the limit $\ell\rightarrow\infty$ is of fundamental importance in the
flow equation scheme since $\tilde u(\infty)$ sets the generated 
mass gap in the spectrum. Via (\ref{rel_u_tildeu}) we can also establish
the following expression for the ``usual'' running coupling $u(\ell)$
in the language of the scaling equations (\ref{flow_beta}).
In the limit of $\ell\rightarrow\infty$ one finds
\beq
u(\ell)=a\,M\,e^\ell \ .
\label{mass2}
\eeq
Since $\Lambda_B\propto B^{-1/2}$ plays the role of an effective 
UV--cutoff generated by the flow equations, this means that the
dimensionless parameter $u(\Lambda_B)$ diverges simply as the mass~$M$ 
divided by the effective cutoff $\Lambda_B$
\beq
u(B)\propto \frac{M}{\Lambda_B} \ ,
\eeq
which allows a simple physical picture of the diverging~$u(B)$ in the
strong--coupling phase.
Again, this does not imply the breakdown of our approach since
$u(B)$ is not the expansion parameter in our approach (see Sect.~IV.D below). 

\subsubsection{Scaling behavior of the mass gap} \noindent
We will now analyze the behavior of the single--particle/hole mass~$M$
in various regimes and compare this with results obtained with other methods.
First we concentrate on the scaling behavior in the limit $u_0\rightarrow 0$
as this can be derived analytically: According to (\ref{mass2}) it amounts 
to finding the scaling invariant 
\beq
I(\ell)=e^{-\ell} f(u(\ell),\beta^2(\ell))
\eeq
with some suitable function $f(u,\beta^2)$ such that 
\beq
\frac{dI(\ell)}{d\ell}=0
\eeq
along the flow generated by (\ref{flow_beta}). Like in the conventional
scaling analysis the mass $M$ is then given by
\beq
M\propto f(u_0,\beta^2_0)/a \ .
\eeq
One finds the same behavior as in two--loop order in Sect.II.B: For example
for fixed $\beta^2_0<8\pi$ and $u_0\rightarrow 0$ one can easily check that
$f(u,\beta^2)=u^{1/(2-\beta^2/4\pi)}$ gives a scaling invariant up to
terms in second order
\beq
\frac{dI(\ell)}{d\ell}=I(\ell)\times O\left( 
\left( \frac{u(\ell)}{2-\beta^2(\ell)/4\pi} 
\right)^2 \right)
\eeq
and therefore 
\beq
M\propto u_0^{1/(2-\beta_0^2/4\pi)}/a
\label{scaling3}
\eeq
in agreement with (\ref{scaling1}). Likewise one also finds
the scaling behavior (\ref{2loop}) and (\ref{scaling2}) since we could
reproduce the perturbative RG--equations (\ref{pert_RG2}) in the vicinity
of $\beta^2=8\pi$ within our flow equation approach. 

Since we find agreement in two--loop order, it is of some interest to
also compare with higher loop calculations~\cite{Amit80}. For simplicity
we focus on the mass gap along $S_-$ in Fig.~\ref{fig_KT}, that is
for $\beta_0^2=8\pi(1-u_0)$ in the limit $u_0\rightarrow 0$. We
write $\beta^2(\ell)=8\pi(1-v(\ell))$ and expand (\ref{flow_alpha_fulla})
and (\ref{flow_alpha_fullb}) up to third order in $u$ and $v$
\bea
\frac{dv}{d\ell}&=&2u^2-4(1+\gamma)u^2 v 
\label{3loopa} \\
\frac{du}{d\ell}&=&2uv-4\left(\frac{\pi}{4}-\frac{\gamma}{2}
-\frac{1}{2}\ln 8\right)\:u^3
\label{3loopb}
\eea
where $\gamma\approx 0.577$ is Euler's constant. It is straightforward
to derive the scaling invariant from these equations and one finds
\beq
M\propto u_0^\tau \exp\left(-\frac{1}{2u_0}\right) \: /a
\eeq
with 
\beq
\tau=\frac{1-\ln 8 +\pi/2}{3} \approx 0.16 \ .
\eeq
This should be compared with the three loop result (\ref{3loop})~\cite{Amit80} 
with the correct exponent $\tau_{\rm RG}=1/2$.\footnote{The 
deviation from the result for $\tau$ in Ref.~\protect\cite{Kehrein99}
occurs because the term in order~$u^3$ in (\ref{flow_alpha_fullb}) was neglected
there.} We see that the present order of our flow
equation expansion is correct up to two loop order and deviates
if compared with three loop RG. This is not surprising since
we have not systematically taken all the terms in order $\tilde u^3(\ell)$ 
into account in the present order of our flow equation scheme and
there are contributions in order~$u^3$
missing on the rhs of (\ref{3loopb}).

\begin{figure}[t]
\begin{center}
\leavevmode
\epsfxsize=10cm
\epsfysize=10cm
\epsfbox{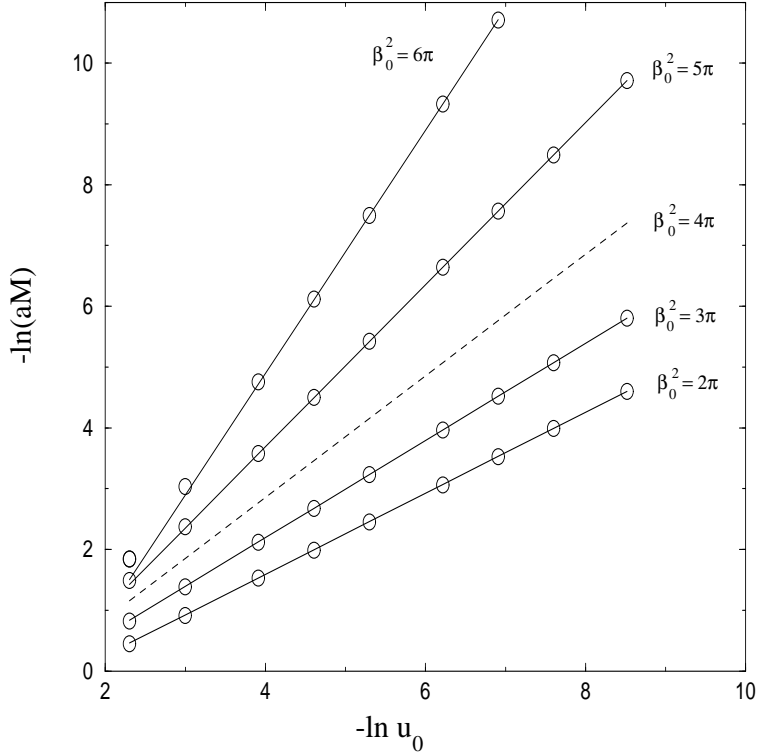}
~\vspace*{0.5cm}~

\caption{Soliton mass $M$ as a function of the coupling constant for various
values of~$\beta_0^2$: The full lines are constrained fits of the power
law behavior $aM\propto u_0^{1/(2-\beta_0^2/4\pi)}$ from the exact
solution Ref.~\protect\cite{Faddeev75} to the flow equation
results (open circles) with the proportionality constant being fitted. The
dashed line is the case $\beta_0^2=4\pi$ where the flow equation
approach agrees trivially (see text).}
\label{fig_mass}
\end{center}
\end{figure} 
In general no closed analytical solution for the set of differential 
equations (\ref{flow_beta}) could be found.
Some numerical solutions of (\ref{flow_beta}) 
are depicted in Fig.~\ref{fig_mass}. Of course the scaling
behavior (\ref{scaling3}) is reproduced by these numerical solutions as
can be seen in Fig.~\ref{fig_mass}. Finally for $\beta_0^2=4\pi$
one can easily prove $M=u_0/a$ exactly using the above flow equations
as should be expected since our scheme becomes exact in this case.

\subsection{Properties of $\mathbf{H(B=\infty)}$} \noindent
Since the flow equation procedure diagonalizes the sine--Gordon Hamiltonian,
we can not only learn something about the mass gap in the spectrum, but
analyze the entire dispersion relation throughout the crossover region.
The final Hamiltonian takes the structure
\beq
H(B=\infty)=H_0+H_{\rm diag}(B=\infty)
\eeq
with
\bea
H_0&=&\int dx\, \left( \frac{1}{2}\Pi^2(x)+\frac{1}{2}\left(\frac{\partial\phi}{\partial x}\right)^2 \right) \\
H_{\rm diag}(B=\infty)&=&\sum_{k>0} \omega(B=\infty;k) \left( P_1^\ph(-k) P_1^\dagger(-k)
+P_1^\dagger(k) P_1^\ph(k)
+ P_2^\dagger(-k) P_2^\ph(-k)+ P_2^\ph(k) P_2^\dagger(k) \right)
\eea
and $P_j(k)$ defined in (\ref{Sjk}). Notice that $[H_0,H_{\rm diag}(B=\infty)]=0$
and $H(B=\infty)|\Omega\rangle=0$. Using (\ref{vertex_exchange}) and (\ref{annihilation}) 
(see also the 
reasoning below (\ref{S2})) and the normalization (\ref{normalization}) one finds the following 
single--particle/hole excitation spectrum of $H(B=\infty)$:
\begin{itemize}
\item Soliton (particle) excitations with excitation energy $E_k$:
\begin{itemize} 
\item[~] $P_1^\dagger(k)|\Omega\rangle$ for $k>0$ 
\item[~] $P_2^\dagger(k)|\Omega\rangle$ for $k<0$
\end{itemize}
\item Antisoliton (hole) excitations with excitation energy $-E_k$:
\begin{itemize} 
\item[~] $P_1^\ph(k)|\Omega\rangle$ for $k<0$ 
\item[~] $P_2^\ph(k)|\Omega\rangle$ for $k>0$
\end{itemize}
\end{itemize}
The dispersion relation $E_k$ is given by
\beq
E_k=|k|+\omega(B=\infty;|k|) \ .
\label{dispersion_1}
\eeq
In order to find $\omega(B=\infty;k)$ we next have to solve the system
of equations (\ref{flow_v}) and (\ref{flow_omega}). A closed analytical
solution has not been possible except for the trivial case
$\beta_0^2=4\pi$ where one reproduces (\ref{dispersion_Thirring}) exactly.
However, we will see below that 
the flow of $\omega(B;k)$ from its initial value~0 to $\omega(B=\infty;k)$
occurs on the $B$--scale~$B_k$~(\ref{def_Bk}) and is negligible for
$B\ll B_k$ or $B\gg B_k$. We can therefore to a good approximation 
replace $\alpha(B)$ and $\tilde u(B)$ in (\ref{flow_omega}) by its values
for $B=B_k$ and consider them as constants. Also one can verify numerically that 
the term in the differential equation proportional to $\sin(\pi\alpha^2)$ changes
the dispersion relation (\ref{dispersion_1}) only in relative order~$1\%$
for $\beta_0^2\geq 4\pi$ (for $|k/M|<5$ it e.g.\ 
affects $\omega(B=\infty;k)$ less then~$2\%$). In order to gain some
first analytical insight we can neglect it. Notice that this approximation
becomes exact in the low--energy limit in the strong--coupling phase
since $\alpha^2(B)\stackrel{B\rightarrow\infty}{\longrightarrow} 1$.
We arrive at ($k>0$)
\bea
\frac{\partial v(B;k)}{\partial B}
&=&-4k^2 v(B;k) -4k\, \omega(B;k) v(B;k)
\label{flow_v2} \\
\frac{\partial\omega(B;k)}{\partial B}
&=&4k v^2(B;k) c_k 
\label{flow_omega2}
\eea
with
\beq
c_k=-\frac{\cos(\pi\alpha^2(B_k))}{\Gamma^2(\alpha^2(B_k))} \: 
\frac{\tilde u^2(B_k)}{a^2}\: |ak|^{2\alpha^2(B_k)-2} \ .
\label{def_ck}
\eeq
Now this approximated system of flow equations can be solved easily.
One finds for $0\geq c_k>-k^2$
\beq
\omega(B;k)=-k+\sqrt{k^2+c_k}\,\coth\left(4k\sqrt{k^2+c_k}\:B\:
+{\rm arccoth}\left(\frac{k}{\sqrt{k^2+c_k}}\right)\right) \ ,
\label{sol_1}
\eeq
and for $c_k\geq 0$
\beq
\omega(B;k)=-k+\sqrt{k^2+c_k}\,\tanh\left(4k\sqrt{k^2+c_k}\:B\:
+{\rm arctanh}\left(\frac{k}{\sqrt{k^2+c_k}}\right)\right) \ .
\label{sol_2}
\eeq
In both cases one obtains
\beq
\omega(B=\infty;k)=-k+\sqrt{k^2+c_k} \ .
\label{final_omega}
\eeq
In the case $c_k \leq -k^2$ the solution for $\omega(B;k)$ diverges 
as $B\rightarrow\infty$. For small initial couplings~$u_0$ this
scenario can according to (\ref{def_ck}) only occur for $\alpha^2(B=0)<1/2$.
This just defines our permissible range of paramters $\beta_0^2\geq 2\pi$ as mentioned above. 
From (\ref{sol_1}) and (\ref{sol_2}) we can also read of the justification for
our above approximation in the system of differential equations: Nearly all
the flow from $\omega(B=0;k)=0$ to $\omega(B=\infty;k)$ occurs on the scale
$B\approx [4k\sqrt{k^2+c_k}]^{-1} \approx B_k$ with $B_k$ from (\ref{def_Bk}).
One can verify numerically that for $\beta_0^2\geq 4\pi$ and $|k|<3\tilde u(\infty)/a$
the solution (\ref{final_omega}) of the 
approximated flow equations agrees to within 20\% with the full numerical
solution of the equations (\ref{flow_v}) and (\ref{flow_omega}) and becomes
exact for $|k|\ll \tilde u(\infty)/a$.

\begin{figure}[t]
\begin{center}
\leavevmode
\epsfxsize=14cm
\epsfysize=14cm
\epsfbox{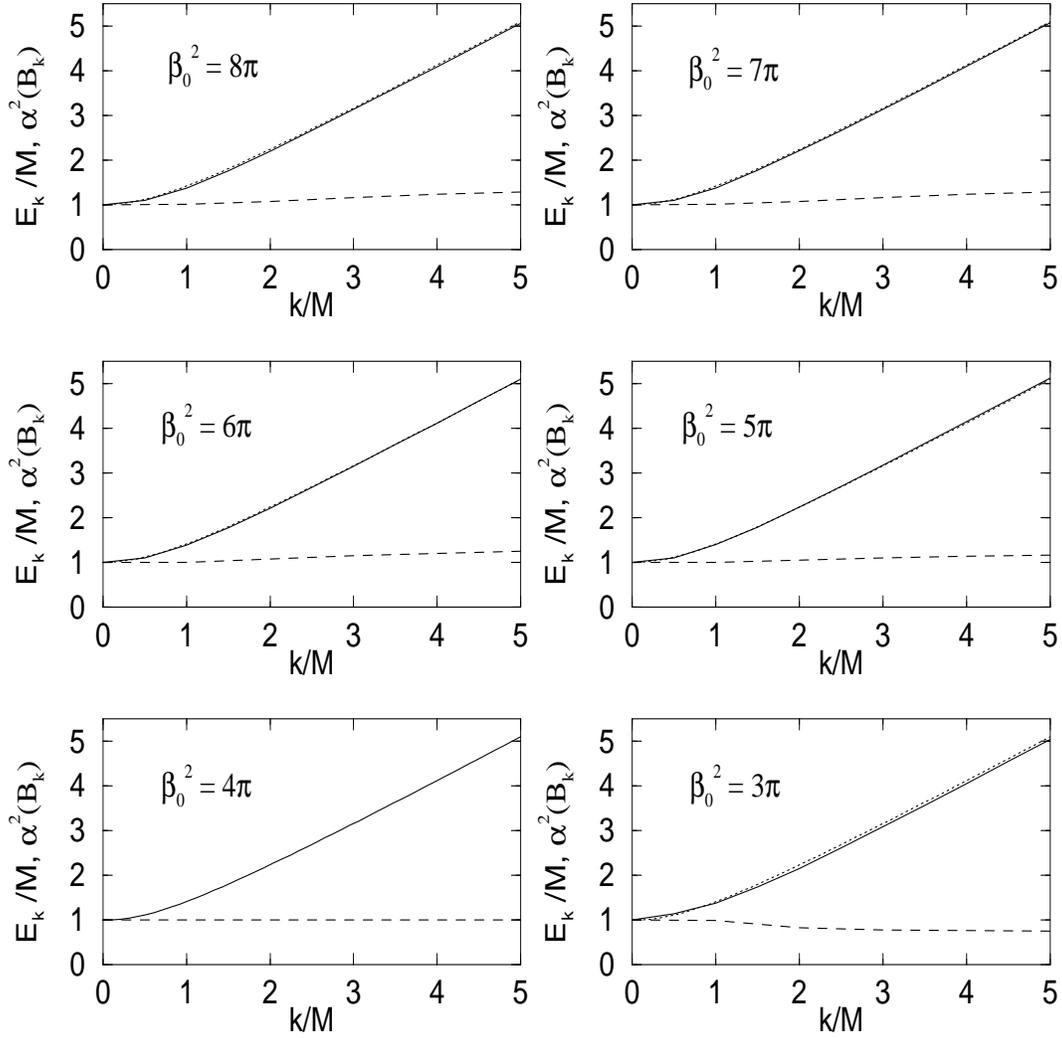}
~\vspace*{0.5cm}~

\caption{Universal scaling limit for the dispersion relation $E_k$ and the
scaling dimension of the soliton excitations obtained by numerical solution
of (\protect\ref{flow_v}) and (\protect\ref{flow_omega}): The full lines
depict $E_k$, the dotted lines $\sqrt{k^2+M^2}$ for comparsion, and the 
dashed lines the scaling dimension $\alpha^2(B_k)$ corresponding to the resp. wavevector.}
\label{fig_dispersion}
\end{center}
\end{figure} 
Putting everything together the dispersion relation is 
\beq
E_k=\sqrt{k^2-\cos(\pi\alpha^2(B_k))\left(\frac{1}{\Gamma(\alpha^2(B_k))}\:
\frac{\tilde u(B_k)}{a}\: |ak|^{\alpha^2(B_k)-1} \right)^2} \ .
\label{dispersion}
\eeq
In the low--energy limit the dispersion relation (\ref{dispersion}) takes 
different forms in the weak-- and strong--coupling phases:
\begin{itemize}
\item Weak--coupling phase:
Here $\alpha^2(B=\infty)\geq 2$ and we find for $|k|\ll \tilde u(\infty)/a$
\beq
E_k=|k|\:\sqrt{1-\cos(\pi\alpha^2(\infty))\left(\frac{1}{\Gamma(\alpha^2(\infty))}\:
\tilde u(\infty) \: |ak|^{\alpha^2(\infty)-2} \right)^2} \ ,
\eeq
that is a gapless spectrum with $E_k=|k|$ for $k\rightarrow 0$.
\item Strong--coupling phase:
Here $\alpha^2(B=\infty)=1$ and we find for $|k|\ll \tilde u(\infty)/a$
\beq
E_k=\sqrt{k^2+\left(\frac{\tilde u(\infty)}{a}\right)^2} \ .
\label{dispersion4}
\eeq
This agrees with the result (\ref{dispersion1}) obtained from the effective
Hamiltonian analysis in Sect.~IV.B.2, that is we find a gapped spectrum with
the mass~$M=\tilde u(\infty)/a$. 
\end{itemize}
One can verify numerically that in the strong--coupling phase for $\beta_0^2\geq 4\pi$
the full dispersion relation obtained by solving
(\ref{flow_v}) and (\ref{flow_omega}) is very accurately described by 
$\sqrt{k^2+M^2}$ even in the crossover region: In the small coupling limit
$|u_0|\ll 1$ there are $\beta_0$--dependent universal corrections in the
crossover region $k=O(M)$ that vanish for $\beta_0^2\rightarrow 4\pi$ 
and reach at most~$3\%$ (for $\beta_0^2=8\pi$).\footnote{Such
small corrections might be expected from the exact results for the 
one--dimensional spin--1/2 XYZ--chain in Ref.~\protect\cite{Johnson73},
where the form (\protect\ref{dispersion4}) holds {\em exactly} also
in the crossover region. However, 
a strict comparison with~\protect\cite{Johnson73} is difficult since there
are nontrivial renormalization subtleties in mapping the continuum
sine--Gordon model to the disrete spin chain (in this context see
e.g.\ Ref.~\protect\cite{Luther75}).} 
The respective scaling forms of the dispersion relation are depicted in 
Fig.~\ref{fig_dispersion}.

\begin{figure}[t]
\begin{center}
\leavevmode
\epsfxsize=10cm
\epsfysize=10cm
\epsfbox{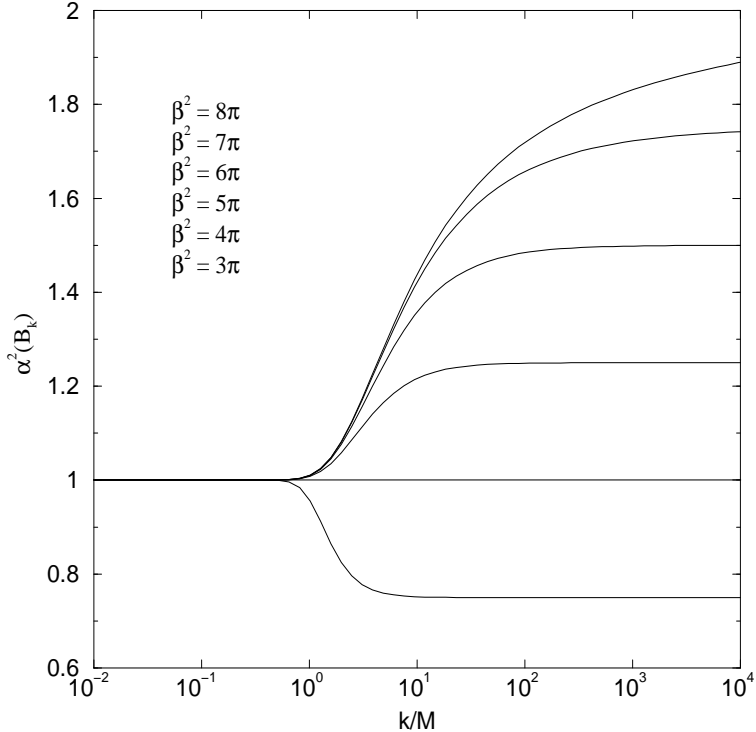}
~\vspace*{0.5cm}~

\caption{Universal curves for the effective scaling dimension $\alpha^2(B_k)$
corresponding to a soliton excitation with wavevector~$k$ in the strong--coupling
phase. The curves are from top to bottom
for $\beta_0^2=8\pi,7\pi,6\pi,5\pi,4\pi,3\pi$.}
\label{fig_effalpha}
\end{center}
\end{figure} 
Notice that according to (\ref{Sjk}) the scaling dimension 
of our single--particle/hole excitations {\em varies} continuously along these 
dispersion curves, see Figs.~\ref{fig_dispersion} and~\ref{fig_effalpha}:
In the strong--coupling phase one finds the initial 
scaling dimension $\alpha(0)$ for excitations with large momenta $|k|\gg M$, and
the low--energy effective Thirring fermions with $\alpha(\infty)=1$ for small
momenta $|k|\ll M$. This is consistent with the exact $S$--matrix results by 
Zamolodchikov~\cite{Zamolodchikov77} discussed in Sect.~II.C. Also notice that our
elementary excitations in this section are described with respect to a transformed
basis since $H(B=\infty)$ and $H(B=0)$ are related by a complicated unitary
transformation.

Finally let us look at the solution of the equations (\ref{flow_v}) and
(\ref{flow_omega}) for initial parameters $\beta_0^2<4\pi$. 
We have seen above that the differential equation for $\omega(B;k)$
leads to divergences for $k\rightarrow 0$ if $\alpha_0^2<1/2$.\footnote{
In fact the $\sin(\pi\alpha^2)$--term in (\ref{flow_omega}) already leads
to IR--problems for $\alpha_0^2<1$. The source of this problem is related
to the breakdown at $\alpha_0^2=1/2$ and can be resolved in a likewise manner
as will be shown in a subsequent publication: The $\sin(\pi\alpha^2)$--terms
turn out to be generally unimportant and we can safely neglect them in our
present discussion also for $\alpha_0^2<1$.} 
In other words the $\cos(\beta_0 \phi(x))$--term becomes too relevant for our
flow equation approach in its present form for $\beta_0^2< 2\pi$.
This is the reason why the parameter space of the sine--Gordon model 
that we can deal with in this paper is {\em restricted to} 
$\beta_0^2\geq 2\pi$. In the interval
$2\pi\leq \beta_0^2 <4\pi$ our approximations become exact
in the limit $\beta_0^2\rightarrow 4\pi$ and, according to the observations
above, eventually break down for $\beta_0^2< 2\pi$. We will investigate the accuracy 
of our approximations in more detail in the following section.
\begin{figure}[t]
\begin{center}
\leavevmode
\epsfxsize=10cm
\epsfysize=10cm
\epsfbox{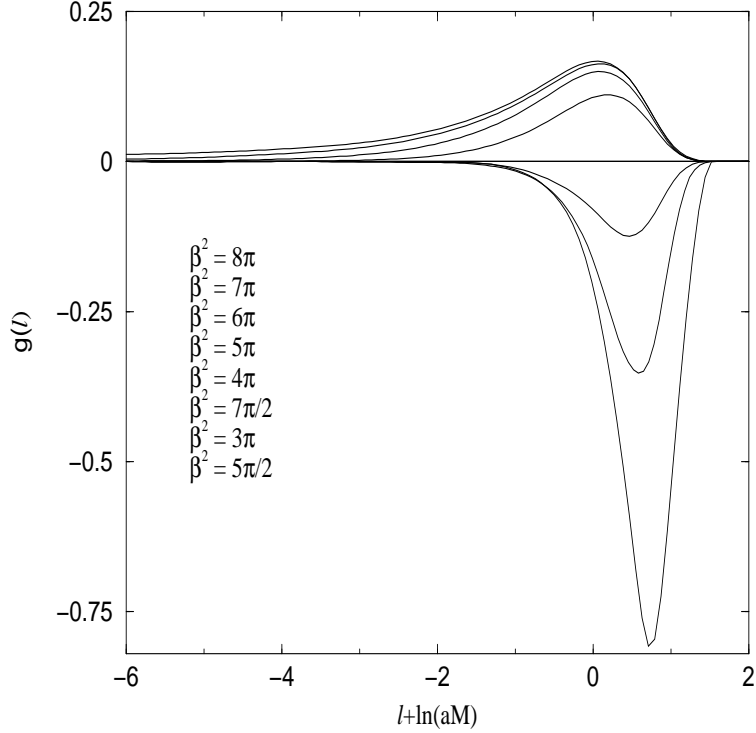}
~\vspace*{0.5cm}~

\caption{Universal curves for the expansion parameter $g(\ell)$ of the flow
equation approach (\protect\ref{def_g}). The curves are from top to bottom
for $\beta_0^2=8\pi,7\pi,6\pi,5\pi,4\pi,7\pi/2,3\pi,5\pi/2$. Notice that
$g(\ell)\equiv 0$ for $\beta_0^2=4\pi$ since then our approach is exact.}
\label{fig_g}
\end{center}
\end{figure} 
\subsection{Approximations and expansion parameter} \noindent
Both during the flow and for the analysis of the final diagonal Hamiltonian
we have neglected the terms in $H_{\rm res}(B)$ in (\ref{full_H}). For any
given $B$--scale these terms are more irrelevant by at least two spatial
derivatives than the interaction term $H_{\rm int}(B)$. Our approximations
are therefore similar to the expansions in renormalization approaches.
To judge the accuracy of our approximations in more detail it is important
to study the prefactors of these terms, that is to find the expansion
parameter of our approach. Since $H_{\rm int}(B)$ is treated as the perturbing
term, we can do this easily by comparing it with $H_0$: E.g.\ from the
dispersion relation (\ref{dispersion}) one sees that on the momentum scale~$k$ 
the effect of $H_{\rm int}(B)$ as compared to $H_0$ is
\bea
\frac{|ak|^{\alpha^2(B_k)}\:\tilde u(B_k)/a}{|k|}
&\propto& \tilde u(B_k) \left( \frac{\sqrt{B_k}}{a} \right)^{2-\alpha^2(B_k)} \nn \\
&=& u(B_k) \ ,
\eea 
where we have used (\ref{tilde_u1}). Not surprisingly the expansion parameter
seems to be the running coupling constant $u(B)$ like in perturbative RG.
According to the systems of differential equations (\ref{flow_beta}) this is
good news in the weak--coupling phase since $u(B)$ decays to zero on small
energy scales ($B\rightarrow\infty$). However, in the strong--coupling phase
$u(B)$ diverges according to (\ref{mass2}). 

Still our method is a systematic approximation even in the strong--coupling
phase since our solution becomes {\em exact} for $\beta^2=4\pi$. As we have seen
in Sect.~III, our system of flow equations closes for $\beta^2=4\pi$
($\alpha^2=1$) and {\em no} higher order interactions are generated during
the flow. Therefore $H_{\rm res}(B)$ vanishes identically on this line. This remarkable
observation is our main difference from perturbative RG. It is in fact even a
trivial observation since according 
to Sect.~II.C the sine--Gordon Hamiltonian becomes equivalent to a noninteracting
Thirring model for $\beta^2=4\pi$. And a quadratic Hamiltonian can easily be
solved exactly with our scheme of unitary transformations. 

Therefore the true expansion parameter of our method is {\em necessarily} some product
of $u(B)$ and $(\alpha^2(B)-1)$. One can check explicitly that the leading terms
in $H_{\rm res}(B)$ contribute like 
\beq
g(B)\stackrel{\rm def}{=}u^2(B)\,(\alpha^2(B)-1) \ .
\label{def_g}
\eeq
In this context also compare our analysis in Sect.~III.D.3 where we have seen in 
(\ref{sigma1sigma2}) that the $R$--term (that we had to treat approximately) vanishes 
linearly like $(\alpha^2-1)$ as $\alpha^2\rightarrow 1$.

This dimensionless combination $g(B)$ is therefore the expansion parameter of our 
scheme\footnote{Also the combination $g(B)\,u(B)$ appears, but this does not make any 
difference for our analysis.} and has to remain small throughout the entire flow in order 
to have a systematic and controllable approximation. 

In order to verify this let us first investigate how $(\alpha^2(B)-1)$ vanishes for 
$B\rightarrow\infty$ in the strong--coupling phase:
We define $\epsilon(\ell)\stackrel{\rm def}{=} \alpha^2(\ell)-1$ and approximate
(\ref{flow_alpha_fulla}) for large~$\ell$ yielding
\beq
\frac{d\epsilon(\ell)}{d\ell}=-\epsilon(\ell)\,a^2 M^2\,e^{2\ell} \ ,
\eeq
which can be solved easily leading to 
\beq
\epsilon(\ell)=\epsilon(\ell_0)\,\exp\left[
-\frac{(aM)^2}{2} \,(e^{2\ell}-e^{2\ell_0}) \right] \ .
\eeq
This very fast decay for $\ell\rightarrow\infty$ avoids a divergence in 
our expansion parameter $g(\ell)$
\beq
g(\ell)\propto (aM)^2 \exp\left[ 2\ell
-\frac{(aM)^2}{2} \,e^{2\ell}\right] 
\stackrel{\ell\rightarrow\infty}{\rightarrow} 0 \ .
\eeq
For the full flow of $g(\ell)$ one finds universal curves 
in the small coupling limit $u_0\rightarrow 0$ that depend on $\beta_0^2$. 
Numerical solutions for these
universal curves are shown in Fig.~\ref{fig_g}. The fact that one finds
universal nonvanishing curves for $\beta_0^2\neq 4\pi$ 
means that there are nonzero corrections to our present approximations
even in the limit $u_0\rightarrow 0$. This is not surprising due to the
strong--coupling nature of our model. A precise statement about
the actual size of the errors is difficult since this depends on unknown prefactors
with which $g(\ell)$ enters. However, it is encouraging to observe that $|g(\ell)|$
remains relatively small throughout the entire flow for all $\beta_0^2\geq 4\pi$.
Higher order corrections can therefore be expected to be small and are systematically 
obtainable in an expansion that takes more and more irrelevant terms into account 
in our flow equation procedure.

Finally, for $\beta_0^2<4\pi$ the maximum in $|g(\ell)|$ grows rapidly as $\beta_0^2$ 
becomes smaller as can be seen in Fig.~\ref{fig_g}. The error of our approximation
therefore becomes larger as $\beta_0^2\rightarrow 0$, which agrees with the observation
in Sect.~IV.C that our present scheme actually breaks down for $\beta_0^2< 2\pi$. This is 
not unexpected since the perturbing $\cos(\beta\phi(x))$--term becomes more and more
relevant for small $\beta_0$. 

\section{Conclusions} \noindent
In this paper we have developed a new approach~\cite{Kehrein99}
for solving the quantum sine--Gordon
model (\ref{sinegordon3})
\beq
H(B=0)=\int dx\, \left( \frac{1}{2}\Pi^2(x)+\frac{1}{2}\left(\frac{\partial\phi}{\partial x}\right)^2
+\frac{u}{\pi a^2} \cos\left[ \beta\int 
d\epsilon\,c(\epsilon)\,\phi(x+\epsilon) \right]\right) 
\label{sinegordon5}
\eeq
by means of infinitesimal unitary transformations as introduced by
Wegner~\cite{Wegner94} and G{\l}azek and Wilson~\cite{GlazekWilson}.
Within an approximation that neglected operators with larger scaling
dimensions (more irrelevant terms) we obtained a flow that
unitarily linked the initial Hamiltonian (\ref{sinegordon5}) to a
diagonal Hamiltonian 
\bea
H(B=\infty)&=&\int dx\, \left( \frac{1}{2}\Pi^2(x)+\frac{1}{2}\left(\frac{\partial\phi}{\partial x}\right)^2 \right) \nn \\
&&+\sum_{k>0} \omega(B=\infty;k) \left( P_1^\ph(-k) P_1^\dagger(-k)
+P_1^\dagger(k) P_1^\ph(k)
+ P_2^\dagger(-k) P_2^\ph(-k)+ P_2^\ph(k) P_2^\dagger(k) \right) \ .
\label{final_H5}
\eea
Here the $P_j(k)$ are soliton and antisoliton creation and annihilation
operators as defined in (\ref{Sjk}). Their dispersion relation $\pm E_k$
was calculated in (\ref{dispersion}). In the small coupling limit $|u|\ll 1$
we found 
$E_k=\sqrt{k^2+M^2}$
(with very small deviations from this form) in the strong--coupling phase, and a gapless spectrum
$E_k=|k|$
in the weak--coupling phase. In the strong--coupling phase, our low--energy
solitons and antisolitons are fermionic (compare Fig.~\ref{fig_effalpha}) 
as known from the exact $S$--matrix
solution~\cite{Zamolodchikov77}. Within our approach their 
mass~$M$ can be obtained by the solution of the flow equations (\ref{flow_beta})
\bea 
\frac{d\beta^{-2}(\ell)}{d\ell}&=&\frac{1}{4\pi\Gamma\big(
-1+\beta^2(\ell)/4\pi\big)}\:u^2(\ell) 
\label{final_flowbeta}  \\
\frac{du}{d\ell}&=&\left(2-\frac{\beta^2(\ell)}{4\pi}\right)\:u(\ell) 
\label{final_flowu}
\eea
via the relation $M=e^{-\ell}\,u(\ell)/a$ in the limit $\ell\rightarrow\infty$.
The above equations have to be integrated from their initial values for $\ell=0$ 
to $\ell=\infty$. Our results for 
the scaling behavior of the mass agree with exact methods and the two--loop
scaling analysis (compare Fig.~\ref{fig_mass}). We also reproduce the phase diagram
Fig.~\ref{fig_KT} of the sine--Gordon model in our approach.

Our equations (\ref{final_flowbeta}) and
(\ref{final_flowu}) are similar to the two--loop RG~equations (\ref{pert_RG}) 
{\em except} for the $\Gamma$--function in the denominator of (\ref{final_flowbeta})
that makes $\beta^2=4\pi$ an {\em attractive strong--coupling fixed point}
of the flow equation method (see Fig.~\ref{fig_flowbeta}).
This is the main difference between our approach and perturbative~RG.
Since the sine--Gordon model for $\beta^2=4\pi$ can be interpreted as a
noninteracting Thirring model, our flow equation procedure becomes exact
at this point and diagonalizes the ensuing quadratic Hamiltonian easily.
The expansion parameter of our approach is therefore {\em not} $u(\ell)$ 
(notice that $u(\ell)$ diverges in the strong--coupling phase as in 
perturbative scaling), but according to (\ref{def_g}) the product 
$g(\ell)=u^2(\ell)\,(-1+\beta^2(\ell)/4\pi)$. 
$g(\ell)$ remains small throughout the entire flow for $\beta^2\geq 4\pi$
(compare Fig.~\ref{fig_g}).
This allows a {\em systematic} improvement of our present approximations
by successively taking terms with larger scaling dimensions into account
in our flow equation procedure. Furthermore, it allows us 
to conclude that our present approximation
already provides a good description of the crossover region, which is notoriously
difficult to study with other techniques. For example, we worked out
the dispersion relation of the single--particle/hole excitations for all
momenta in Fig.~\ref{fig_dispersion}. Notice that higher order terms in
our expansion {\em cannot} endanger the stability of the strong--coupling 
fixed point.

As can be deduced from Fig.~\ref{fig_g}, our approximations become less
accurate for $\beta^2<4\pi$ and eventually our present approach breaks down
for $\beta^2< 2\pi$ (Sect.~IV.C). In addition, the bound states present in the
spectrum for $\beta^2<4\pi$ according to the exact solution
are absent in our solution. These bound states will be generated by
interactions in $H_{\rm res}$ that are not included in our present approximation. 
Work on these issues for $\beta^2<4\pi$ is in progress.

To summarize, we have obtained an explicit approximate relation between the 
strong--coupling problem (\ref{sinegordon5}) and its diagonalized form (\ref{final_H5})
without using the integrable structure of the model. The method
presented here provides a theoretical tool that is capable
of achieving this in a systematic expansion throughout the crossover region.
We have been able to carry out this
program completely for $\beta^2\geq 4\pi$, and obtained first results 
for $\beta^2<4\pi$ (where more work remains to be done e.g.\ regarding the
bound states). Our approach is conceptually simple since a small parameter is 
identified and used as an expansion parameter. It is therefore
possible to study nonintegrable perturbations with our
approach, and the calculation of correlation functions also seems feasible.
Finally, there are various
other onedimensional strong--coupling problems, as for example the Kondo model,
where the present approach should be useful~\cite{Hofstetter00}.

The author acknowledges many valuable discussions with W.~Hofstetter and D.~S.~Fisher.
This work was supported by the Deutsche Forschungsgemeinschaft (DFG), by
the SFB~484 of the DFG and by the National Science Foundation (NSF) under 
grants DMR~9630064, DMR~9976621 and DMR~9981283.
\appendix
\section*{Properties of vertex operators}
\subsection{Operator product expansion} \noindent
This Appendix compiles some important properties of vertex operators.
For a review of these properties see also Ref.~\cite{Delft98}.

We first want to establish a relation between the normal--ordered and the 
non--normal--ordered vertex operators. By definition
\bea
V_1(\alpha;x)&\stackrel{\rm def}{=}&\: :\exp\left(\alpha \sum_{p\neq 0} \frac{\sqrt{|p|}}{p}\,
e^{-\frac{a}{2}|p|-ipx} \sigma_1(p) \right): \nn \\
&=&\exp\left(\alpha \sum_{p>0} \frac{\sqrt{|p|}}{p}\,
e^{-\frac{a}{2}|p|-ipx} \sigma_1(p) \right)\:
\exp\left(\alpha \sum_{p<0} \frac{\sqrt{|p|}}{p}\,
e^{-\frac{a}{2}|p|-ipx} \sigma_1(p) \right) 
\eea
and next we can use the formula
\beq
e^A\,e^B=e^{C/2}\,e^{A+B}
\eeq
with $C=[A,B]$ since $C$ is a number and commutes with both $A$ and $B$:
\bea
C&=&\left[\:\alpha \sum_{p>0} \frac{\sqrt{|p|}}{p}\,
e^{-\frac{a}{2}|p|-ipx} \sigma_1(p),
\alpha \sum_{q<0} \frac{\sqrt{|q|}}{q}\,
e^{-\frac{a}{2}|q|-iqx} \sigma_1(q)\:\right] \nn \\
&=&\alpha^2 \sum_{p>0} \, \frac{e^{-a|p|}}{p} \ .
\eea
The sum over~$q$ yields
\beq
C=-\alpha^2 \ln\left(1-e^{-2\pi a/L}\right)
\eeq
and in the thermodynamic limit $L\rightarrow\infty$
\beq
C=-\alpha^2  \ln\left(\frac{2\pi a}{L}\right) \ .
\eeq
Therefore
\bea
V_j(\alpha;x)&\stackrel{\rm def}{=}&
\: :\exp\left(\pm\alpha \sum_{p\neq 0} \frac{\sqrt{|p|}}{p}\,
e^{-\frac{a}{2}|p|-ipx} \sigma_j(p) \right): \nn \\
&=&\left(\frac{L}{2\pi a}\right)^{\alpha^2/2}\:
\exp\left(\pm\alpha \sum_{p\neq 0} \frac{\sqrt{|p|}}{p}\,
e^{-\frac{a}{2}|p|-ipx} \sigma_j(p) \right) \ .
\label{rel_normalordering}
\eea
An important property of the Fourier--transformed vertex operators (\ref{fourier})
is their action on the vacuum
\beq
V_1(\alpha;-k)|\Omega\rangle=V_1(-\alpha;k)|\Omega\rangle
=V_2(\alpha;k)|\Omega\rangle=
V_2(-\alpha;-k)|\Omega\rangle=0 \quad \forall k>0 \ .
\label{prop_S}
\eeq
This is shown easily, e.g.
\bea
V_1(-\alpha;k)|\Omega\rangle&=&
\frac{1}{2\pi} \int dx \: e^{-ikx} V_1(-\alpha;x)|\Omega\rangle \nn \\
&=&\frac{1}{2\pi} \int dx \: e^{-ikx}\:
\exp\left(\alpha \sum_{p>0} \frac{\sqrt{|p|}}{p}\,
e^{-\frac{a}{2}|p|-ipx} \sigma_1(p) \right)\:|\Omega\rangle \ .
\eea
One expands the second exponential and arrives at terms with the
structure
\beq
\int dx\: e^{-ix(k+p_1+\ldots+p_n)} = 0
\eeq
since $k,p_1,\ldots,p_n>0$. Therefore
\beq
V_1(-\alpha;k)|\Omega\rangle=0
\eeq
for $k>0$ and the analysis for the rest of
(\ref{prop_S}) proceeds likewise. 

Next we want to want to evaluate expectation values for products of vertex operators.
We look at the product
\bea
V_1(\alpha;x)V_1(-\alpha;y)&=&
\exp\left(\alpha\sum_{p>0}\frac{\exp(-\frac{a}{2}|p|-ipx)}{\sqrt{|p|}}\sigma_1(p)\right)\;
\exp\left(-\alpha\sum_{p<0}\frac{\exp(-\frac{a}{2}|p|-ipx)}{\sqrt{|p|}}\sigma_1(p)\right)\nn\\
&&\times\exp\left(-\alpha\sum_{q>0}\frac{\exp(-\frac{a}{2}|q|-iqy)}{\sqrt{|q|}}\sigma_1(q)\right)\;
\exp\left(\alpha\sum_{q<0}\frac{\exp(-\frac{a}{2}|q|-iqy)}{\sqrt{|q|}}\sigma_1(q)\right) 
\eea
and commute the second and third exponentials using the formula
\beq
e^A e^B = e^C e^B e^A
\eeq
with $C=[A,B]$ since again $C$ is a number and commutes with both $A$ and $B$:
\bea
C&=&\left[-\alpha\sum_{p<0}\frac{\exp(-\frac{a}{2}|p|-ipx)}{\sqrt{|p|}}\sigma_1(p),
-\alpha\sum_{q>0}\frac{\exp(-\frac{a}{2}|q|-iqy)}{\sqrt{|q|}}\sigma_1(q)\right] \nn \\
&=& \alpha^2 \sum_{q>0} \frac{\exp(-a|q|+iq(x-y))}{q} \nn \\
&=&-\alpha^2 \ln\left(1-e^{i\frac{2\pi}{L}(x-y+ia)}\right) \ .
\label{cnumber}
\eea
In the thermodynamic limit $L\rightarrow\infty$ this gives
\beq
e^C=\left(\frac{L/2\pi}{i(y-x)+a}\right)^{\alpha^2} \ .
\eeq
Therefore
\bea
V_1(\alpha;x)V_1(-\alpha;y)&=&\left(\frac{L/2\pi}{i(y-x)+a}\right)^{\alpha^2}\;
\exp\left(\alpha\sum_{p>0}\frac{\exp(-\frac{a}{2}|p|-ipx)}{\sqrt{|p|}}\sigma_1(p)\right)\;
\exp\left(-\alpha\sum_{q>0}\frac{\exp(-\frac{a}{2}|q|-iqy)}{\sqrt{|q|}}\sigma_1(q)\right)
\nn\\
&&\times\exp\left(-\alpha\sum_{p<0}\frac{\exp(-\frac{a}{2}|p|-ipx)}{\sqrt{|p|}}\sigma_1(p)\right)\;
\exp\left(\alpha\sum_{q<0}\frac{\exp(-\frac{a}{2}|q|-iqy)}{\sqrt{|q|}}\sigma_1(q)\right) \ .
\label{app1}
\eea
From this we obtain the expectation value
\beq
\langle V_1(\alpha;x)V_1(-\alpha;y) \rangle
=\left(\frac{L/2\pi}{i(y-x)+a}\right)^{\alpha^2}  \ .
\label{vertex_expectationvalue1}
\eeq
A similar calculation can be done for $V_2(\alpha;x)$ and the only difference is
the exchange of $x$ and $y$ in the denominator
\beq
\langle V_2(\alpha;x)V_2(-\alpha;y) \rangle
=\left(\frac{L/2\pi}{i(x-y)+a}\right)^{\alpha^2}  \ .
\label{vertex_expectationvalue2}
\eeq
The operator product expansion (OPE) for vertex operators can be
deduced from (\ref{app1})
\bea
V_1(\alpha;x)V_1(-\alpha;y)&=&\left(\frac{L/2\pi}{i(y-x)+a}\right)^{\alpha^2}\;
\exp\left(\alpha\sum_{p>0}\frac{\exp(-\frac{a}{2}|p|)}{\sqrt{|p|}}
(e^{-ipx}-e^{-ipy})\sigma_1(p)\right)
\nn \\
&&\qquad\qquad\times\exp\left(-\alpha\sum_{p<0}\frac{\exp(-\frac{a}{2}|p|)}{\sqrt{|p|}}
(e^{-ipx}-e^{-ipy})\sigma_1(p)\right) \nn \\
&=&\left(\frac{L/2\pi}{i(y-x)+a}\right)^{\alpha^2}\;
\exp\left(-i\alpha (x-y)\sum_{p>0} e^{-\frac{a}{2}|p|-ipx} \sqrt{|p|}\: \sigma_1(p)
+O((x-y)^2) \right) \nn \\
&& \qquad\qquad\times\exp\left(-i\alpha (x-y)\sum_{p<0} e^{-\frac{a}{2}|p|-ipx} 
\sqrt{|p|}\:  \sigma_1(p)
+O((x-y)^2) \right) \nn \\
&=&\left(\frac{L/2\pi}{i(y-x)+a}\right)^{\alpha^2}\;
\left(1+i\alpha (y-x) \sum_{p\neq 0} e^{-\frac{a}{2}|p|-ipx} \sqrt{|p|} \: \sigma_1(p)
+O((x-y)^2) \right) \ .
\label{vertex_OPE1}
\eea
Higher order terms in the OPE can easily be deduced using the above scheme.
These terms are less singular as $x\rightarrow y$, or, in the language of renormalization
theory, they have a larger scaling dimension (are more irrelevant) and can
be expressed as spatial derivatives of the bosonic field. 

A similar calculation for $V_2(\alpha;x)$ gives
\beq
V_2(\alpha;x)V_2(-\alpha;y)
=\left(\frac{L/2\pi}{i(x-y)+a}\right)^{\alpha^2}\;
\left(1+i\alpha (x-y) \sum_{p\neq 0} e^{-\frac{a}{2}|p|-ipx} \sqrt{|p|} \: \sigma_2(p)
+O((x-y)^2) \right) \ ,
\label{vertex_OPE2}
\eeq
again the only difference is the exchange of $x$ and $y$.

It is well--known that for $\alpha=\pm 1$ the vertex operators describe
fermion creation and annihilation operators. This can be checked easily
by using the OPE (\ref{vertex_OPE1}) 
in the anticommutator for the special case $\alpha=1$
\bea
\{V_1(1;x),V_1(-1;y)\}
&=&\left(\frac{L/2\pi}{i(y-x)+a}+\frac{L/2\pi}{i(x-y)+a}\right)\:
\left(1+i\alpha (y-x) \sum_{p\neq 0} e^{-\frac{a}{2}|p|-ipx} \sqrt{|p|} \: \sigma_1(p)
+O((x-y)^2) \right) \nn \\
&\stackrel{a\rightarrow 0}{=}&
L\,\delta(x-y)\:\left(1+i\alpha (y-x) \sum_{p\neq 0} e^{-\frac{a}{2}|p|-ipx} 
\sqrt{|p|} \: \sigma_1(p)
+O((x-y)^2) \right) \nn \\
&=&L\,\delta(x-y) 
\label{vertex_fermion}
\eea
in the limit $a\rightarrow 0$. All higher order terms in the OPE vanish in this limit. 
Likewise one finds
\beq
\{V_1(1;x),V_1(1;y)\}=\{V_1(-1;x),V_1(-1;y)\}
\stackrel{a\rightarrow 0}{=} 0
\eeq
and the same relations for $V_2(\pm 1;x)$.

\subsection{Exchange relations} \noindent
Let us look at the commutation relation of vertex operators
in momentum space. These are worked out in the following for
general~$\alpha$. For simplicity we only consider $V_1(\pm\alpha;x)$, the 
calculation for $V_2(\pm\alpha;x)$ proceeds along similar lines. 
From the definition of the vertex operators one easily verifies the following relation
\beq
V_1(-\alpha;x) V_1(\alpha;y)=\frac{ [-i(y-x)+a]^{\alpha^2} }{ [i(y-x)+a]^{\alpha^2} }
V_1(\alpha;y) V_1(-\alpha;x) \ .
\label{reflection}
\eeq
For large distances $|x-y|\gg a$ the coefficient can be well approximated by
\beq
\cos(\pi\alpha^2)+i{\rm sgn}(x-y) \sin(\pi\alpha^2) \ ,
\eeq
whereas for small distances it becomes equal to 1. Now for small distances the operator
product expansion for the two vertex operators can be used and it is then possible
to write generally for all $x-y$   
\beq
V_1(-\alpha;x) V_1(\alpha;y)-{\rm OPE}(x\rightarrow y) 
=\left(\cos(\pi\alpha^2)+i{\rm sgn}(x-y) \sin(\pi\alpha^2)\right) \:
\left( V_1(\alpha;y) V_1(-\alpha;x)- {\rm OPE}(x\rightarrow y) \right) 
\label{approx_vertex}
\eeq
For our purposes here it will be sufficient to look only at 
the leading $c$--number term in the OPE (higher orders can easily be
taken into account if necessary). This is equivalent to subtracting
the ground state expectation value on both sides (\ref{def_star}).
Notice that our relation ``closes''
\bea
*V_1(-\alpha;x) V_1(\alpha;y)* 
&=&\left(\cos(\pi\alpha^2)+i{\rm sgn}(x-y) \sin(\pi\alpha^2)\right)
\: *V_1(\alpha;y) V_1(-\alpha;x)* \nn \\
&=&\left(\cos(\pi\alpha^2)+i{\rm sgn}(x-y) \sin(\pi\alpha^2)\right) \nn \\
&&\times\left(\cos(\pi\alpha^2)+i{\rm sgn}(y-x) \sin(\pi\alpha^2)\right)\;
*V_1(-\alpha;x) V_1(\alpha;y)* \nn \\
&=& *V_1(-\alpha;x) V_1(\alpha;y)*
\label{close}
\eea
In terms of Fourier components (\ref{fourier}) this reads ($\alpha>0$)
\bea
*V_1(-\alpha;k_1)V_1(\alpha;k_2)*  
&=&\frac{1}{2\pi}\sum_p  *V_1(\alpha;k_2+p)V_1(-\alpha;k_1+p)*\;
\int dx\,e^{ipx} \left(\cos(\pi\alpha^2)+i\:{\rm sgn}(x)\:
\sin(\pi\alpha^2) \right) \nn \\
&=&\cos(\pi\alpha^2) *V_1(\alpha;k_2)V_1(-\alpha;k_1)*  \nn \\
&&+\sin(\pi\alpha^2)\: \frac{2}{\pi} \sum_{n=0}^\infty
\frac{1}{2n+1}  \;
\Big[ *V_1\left(\alpha;k_2-\frac{2\pi}{L}(2n+1)\right) 
V_1\left(-\alpha;k_1-\frac{2\pi}{L}(2n+1)\right)* \nn \\
&&\qquad\qquad\qquad\qquad\qquad -*V_1\left(\alpha;k_2+\frac{2\pi}{L}(2n+1)\right) 
V_1\left(-\alpha;k_1+\frac{2\pi}{L}(2n+1)\right)* \Big] \nn \\
&=&\cos(\pi\alpha^2) *V_1(\alpha;k_2)V_1(-\alpha;k_1)* \nn \\ 
&&-\frac{1}{\pi}\,\sin(\pi\alpha^2)\: {\rm Re} \int dq\: 
\frac{1}{q+i\epsilon} \: *V_1(\alpha;k_2+q)V_1(-\alpha;k_1+q)* \nn \\
&=& -\frac{1}{\pi}\:{\rm Im}\left[ e^{i\pi\alpha^2}
\int dq\: \frac{1}{q+i\epsilon} \: *V_1(\alpha;k_2+q)V_1(-\alpha;k_1+q)* \right]
\label{vertex_exchange}
\eea
in the thermodynamic limit with $\lim_{\epsilon\downarrow 0}$ being understood. 
Likewise
\beq
*V_1(\alpha;k_1)V_1(-\alpha;k_2)*
=-\frac{1}{\pi}\:{\rm Im}\left[ e^{-i\pi\alpha^2}
\int dq\: \frac{1}{q+i\epsilon} \: *V_1(-\alpha;k_2+q)V_1(\alpha;k_1+q)*\right] \ .
\eeq
Notice that these relations close in $k$--space in the same
sense as (\ref{close}). They are only ``simple'' for integer~$\alpha^2$
when the vertex operators behave as bosons or fermions: Then the 
term proportional to $\sin(\pi\alpha^2)$ with its summation over
wavevectors vanishes.


\end{document}